\documentclass[12pt,twoside]{article}   
\usepackage[super,square,numbers,sort&compress]{natbib}

\usepackage{fancyhdr}		

\usepackage[section]{placeins}   %
\usepackage{graphicx}
\usepackage[numbers]{natbib} \setcitestyle{open={},close={}}

\makeatletter \renewcommand\@biblabel[1]{$^{#1}$} \makeatother
\newcommand{\cen}[1]{\begin{center} #1 \end{center}}

\usepackage[mathlines]{lineno}

\usepackage{hyperref}
\hypersetup{ colorlinks,
	citecolor=blue,
	filecolor=blue,
	linkcolor=black,
	urlcolor=blue
}

\usepackage{xcolor}
\definecolor{gray}{rgb}{0.6,0.6,0.6}
\definecolor{red}{rgb}{0.85,0,0}
\definecolor{green}{rgb}{0,0.85,0}
\definecolor{blue}{rgb}{0,0,0.85}
\definecolor{beige}{rgb}{0.92,0.87,0.78}

\usepackage[all]{hypcap}    
\usepackage{amsmath,graphicx}
\usepackage{graphicx} 
\usepackage{bm}
\usepackage{amsmath}
\usepackage{amssymb}
\usepackage{overpic}
\usepackage{bbm}
\usepackage{dsfont}
\usepackage{caption}
\usepackage{rotating}
\usepackage{algorithm} 
\usepackage{algorithmic} 
\usepackage{array}
\usepackage{booktabs}
\usepackage{multirow} 
\usepackage{caption}
\usepackage{graphicx}
\usepackage{subfigure}
\usepackage{float}
\usepackage{tikz}
\usetikzlibrary{spy} 
\usepackage{color}
\usepackage{wrapfig}
\usepackage[absolute,overlay]{textpos}
\usepackage{xr}

\usepackage{diagbox}
\usepackage{hyperref}

\usepackage{subfloat}

\usepackage[capitalize,noabbrev]{cleveref}

\crefformat{equation}{ (#2#1#3) }
\Crefname{figure}{Fig.}{Figures}
\crefrangelabelformat{figure}{#3#1#4--#5{#2}#6}

\usepackage[compact]{titlesec}
\titlespacing{\section}{0pt}{*0}{*0}
\titlespacing{\subsection}{0pt}{*0}{*0} 
\titlespacing{\subsubsection}{0pt}{*0}{*0}

\usepackage{tikz}
\usetikzlibrary{positioning}
\tikzset{>=stealth}

\newcommand{\x}{\mathbf x}
\newcommand{\y}{\mathbf y}
\newcommand{\A}{\mathbf A}
\newcommand{\omg}{\mathbf \Omega}
\newcommand{\tabincell}[2]{
	\begin{tabular}{@{}#1@{}}#2\end{tabular}
} 
\renewcommand{\P}{\mathbf P}
\newcommand{\z}{\mathbf z}

\newcommand{\Z}{\mathbf Z}

\newcommand{\W}{\mathbf W}

\newcommand{\I}{\mathbf I}
\newcommand{\D}{\mathbf D}

\newcommand{\dd}		{\mathbf{d}}
\newcommand{\ee}		{\mathbf{e}}
\newcommand{\aaa}		{\mathbf{a}}

\newcommand{\E}	     {\mathbf{E}}
\newcommand{\X}		{\mathbf{X}}
\newcommand{\G}		{\mathrm{G}}
\newcommand{\alp}	 {\bm{\alpha}}
\newcommand{\alpr}	 {\bm{\alpha 1'}}
\newcommand{\TT}	 {\mathcal{T}}

\renewcommand{\log}{\textup{log}}

\newlength\myindent
\newcommand\bindent{%
	\begingroup
	\setlength{\itemindent}{\myindent}
	\addtolength{\algorithmicindent}{\myindent}
}
\newcommand\eindent{\endgroup}

\newlength\mmindent
\newcommand*{\RED}[1]{{\color{red}#1}}

\begin{document}
	\bibliographystyle{unsrt}
	
	\cen{\sf {\Large {\bfseries  An Improved Iterative Neural Network for High-Quality Image-Domain Material Decomposition in Dual-Energy CT} \\  
			\vspace*{10mm}
			Zhipeng Li$^1$,~Yong Long$^1$,~Il Yong Chun$^2$} \\
		$^{1}$ University of Michigan - Shanghai Jiao Tong University Joint Institute,\\
		Shanghai Jiao Tong University, Shanghai 200240, China \\
		$^{2}$School of Electronic and Electrical Engineering, Sungkyunkwan University,\\ Suwon, Gyeonggi 16419, Republic of Korea 
		\vspace{5mm}\\
		Version typeset \today\\
	}
	\vspace{-0.1in}
	\pagenumbering{roman}
	\setcounter{page}{1}
	\pagestyle{plain}
	Yong Long and Il Yong Chun are corresponding authors. \\Email: yong.long@sjtu.edu.cn; iychun@skku.edu  \\
	
	\renewcommand{\thefootnote}{}
	\footnotetext{This paper has supplementary material. The prefix ``S" indicates the numbers in section, equation, and figure in the supplementary material.}
	
	\vspace{-0.3in}
	\begin{abstract}
		\noindent {\bf Purpose:}  Dual-energy computed tomography (DECT) has been widely used in many applications that need material decomposition.
		Image-domain methods directly decompose material images from high- and low-energy attenuation images, and thus, are susceptible to noise and artifacts on attenuation images.
		The purpose of this study is to develop an improved iterative neural network (INN) for high-quality image-domain material decomposition in DECT, and to study its properties.
		\\
		{\bf Methods:} 
		\RED{
		}
		We propose a new INN architecture for DECT material decomposition.
		The proposed INN architecture uses distinct cross-material convolutional neural network (CNN) in image refining modules, and uses image decomposition physics in image reconstruction modules.
		The distinct cross-material CNN refiners incorporate distinct encoding-decoding filters and cross-material model that captures correlations between different materials.
		We study the distinct cross-material CNN refiner with patch-based reformulation and tight-frame condition. \\
		{\bf Results:} Numerical experiments with extended cardiac-torso phantom and clinical data show that the proposed INN significantly improves the image quality over
		several image-domain material decomposition methods, including a conventional model-based image decomposition (MBID) method using an edge-preserving regularizer, a recent MBID
		method using pre-learned material-wise sparsifying transforms, and a noniterative deep CNN method.
		Our study with patch-based reformulations reveals that learned filters of distinct cross-material CNN refiners can approximately satisfy the tight-frame condition.
		\\
		{\bf Conclusions:} 
		The proposed INN architecture achieves high-quality material decompositions using iteration-wise refiners that exploit cross-material properties between different material images with distinct encoding-decoding filters.
		Our tight-frame study implies that cross-material CNN refiners in the proposed INN architecture are useful for noise suppression and signal restoration.
		\\		
	\end{abstract}

	\newpage     

	\tableofcontents
	
	\newpage
	
	\setlength{\baselineskip}{0.7cm}      
	
	\pagenumbering{arabic}
	\setcounter{page}{1}
	\pagestyle{fancy}
	
	\section{Introduction}
	Dual-energy CT (DECT) has been increasingly used in many clinical and industrial applications, 
	including kidney stone characterization~\cite{hokamp2020dose},
	iodine quantification~\cite{jacobsen2019dual,Li2013Iodine},
	security inspection~\cite{Liu2010Feasibility,DECTSecurity}, and nondestructive testing~\cite{Engler1990Review}.
	Compared to conventional single-energy X-ray CT, DECT provides two sets of attenuation measurements at high and low energies. 
	Because the linear attenuation coefficient is material and energy dependent, DECT can characterize different constituent materials in a mixture, known as material decomposition~\cite{mendonca:14:afm}.
	Decomposed material images provide the elemental material compositions of the imaged object.
	Researchers have been studying material decomposition or reconstruction with spectral CT~\cite{wu:19} and photon-counting CT~\cite{wu:20:tim} that can simultaneously acquire more than two spectral measurements. 

	\subsection{Literature Review}
	\label{sec:background}
	Model-based image decomposition (MBID) methods incorporate material composition physics, statistical model of measurements, and some prior information of unknown material images. 
	Existing MBID methods for DECT can be classified into direct (projection-to-image domain)~\cite{long:14:mmd}, projection-domain~\cite{noh:09:ssr}, and image-domain~\cite{niu2014iterative} decompositions.
	Direct decomposition methods perform image decomposition and reconstruction simultaneously, and generate material images directly from collected high and low energy measurements.
	This type of methods can reduce the cross-talk and beam-hardening artifacts by using an accurate forward model of the DECT system along with priors.
	However, direct decomposition algorithms need large computational costs, 
	because at each iteration, they apply computationally expensive forward and backward projection operators.
	Projection-domain methods first decompose high- and low-energy sinograms into sinograms of materials, followed by an image reconstruction method such as filtered back projection (FBP) to obtain material images.
	Although above two types of methods improve the decomposition accuracy compared to image-domain methods, they usually require accurate system calibrations that use nonlinear models~\cite{MM:11:accu,MarinState}. 
	In addition, those methods require sinograms or pre-log measurements that are in general not readily available from commercial CT scanners.
	Image-domain methods do not require projection operators and decompose readily available reconstructed high- and low-energy attenuation images into material images, and are more computationally efficient than direct and projection-domain decomposition methods.
	However, image-domain methods lack complete DECT imaging model. 
	This may increase noise and artifacts in decomposed material images.

	To improve image-domain DECT material decomposition methods, incorporating appropriate prior knowledge or regularizer into decomposition algorithms is critical.
	Many MBID methods have been proposed from this perspective. 
	Niu \textit{et al.}~\cite{niu2014iterative} proposed an iterative decomposition method that incorporates the noise variance of two attenuation images into the least-squares data-fit term.
	This better suppressed noise and artifacts on decomposed material images than a simple direct matrix inversion method.
	Xue \textit{et al.}~\cite{xue:2017:statistical} proposed an MBID method that uses the weighted least-squares data-fit model~\cite{niu2014iterative} and an edge-preserving (EP) hyperbolar regularizer—called DECT-EP.
	Recently, there has been growing interest in data-driven methods such as MBID using pre-learned prior operators.
	Examples include learned synthesis operator/dictionary~\cite{CDL:18:ilyong,wu:19:dlimd} and analysis operator/transform~\cite{CAOL:19:il,li2018image}.
	Dictionary learning has been applied to image-domain DECT material decomposition~\cite{wu:19:dlimd} and improved image decomposition compared to non-adaptive MBID methods. 
	We proposed a data-driven method DECT-ST~\cite{li2018image} that uses two pre-learned sparsifying transforms (ST) in a prior model to better sparsify the two different materials, and improved the image decomposition accuracy.
	We also proposed a clustering based cross-material method~\cite{li:19:immmd} that assumes correlations between different materials, and followed by a generalized mixed material method~\cite{li:19:multra} that considers both individual properties (e.g., different materials have different densities and structures) and correlations of different material images.
	
	In the past few years, deep regression neural network (NN) methods have been gaining popularity in medical imaging applications, for example, CT image denoising~\cite{wu2017cascaded,Jin2017DeepCN}. 
	Several deep convolutional NN (dCNN) methods have also been proposed for image-domain DECT material decomposition.
	Liao \textit{et al.}~\cite{YuCNN} proposed a cascaded dCNN method to obtain a material image from a single energy attenuation image. 
	The first dCNN roughly maps a single attenuation image to a material image, followed by the other dCNN maps the material image to a high-quality material image. 
	A dCNN method with two input and output channels that directly maps from two high- and low-energy attenuation images to two material images has also been proposed~\cite{Xu:2018:ImgCNN} .
	Different from the first dCNN used in aforementioned cascaded dCNN method~\cite{YuCNN} that obtains two material images individually,
	butterfly network~\cite{Niu:18:butterfly} decomposes material images with additional CNNs between two attenuation images to perform information exchange.
	Clark \textit{et al.}~\cite{clark2018multi} investigated the conventional U-Net architecture for image-domain multi-material decomposition. 
	However, the aforementioned methods have the high NN complexity that can increase the overfitting risk particularly when limited training samples are available.

	An alternative approach is a so-called iterative NN (INN), which has been successfully applied to diverse imaging problems~\cite{chun:18:dbn,Chun&etal:18Allerton,bcdnet:19:il,bcdnet:19:Hongki,momen:19:il,momen:20:Ye,ADMMNet}.
	This approach incorporates iteration-wise image refining NNs into block-wise model-based image reconstruction algorithm.
	INN improves generalization capability compared to noniterative deep NN by balancing imaging physics and prior information estimated via refining CNNs,  
	particularly when training samples are limited~\cite{bcdnet:19:il,bcdnet:19:Hongki}.
	ADMM-Net is a pioneer INN architecture developed by unrolling the alternating direction method of multipliers (ADMM) model-based image reconstruction (MBIR) algorithm~\cite{ADMMNet}; 
	it has been succesfully applied to highly-undersampled MRI~\cite{ADMMNet}, low-dose CT~\cite{bcdnet:19:il}, etc.
	BCD-Net is an INN architecture that generalizes the block coordinate descent (BCD) MBIR algorithm using learned convolutional regularizers, while showing better performance over ADMM-Net~\cite{bcdnet:19:il,momen:19:il}.
	Its original work~\cite{chun:18:dbn} uses the identical encoding-decoding architecture, i.e., each filter in decoder is a rotated version of that in encoder, and was successfully applied to highly-undersampled MRI (using single coil).
	Subsequent works~\cite{bcdnet:19:il,bcdnet:19:Hongki} use the distinct encoding-decoding architecture for BCD-Net, and successfully applied modified BCD-Net to low-dose CT and low-count PET reconstruction.
	The Momentum-Net architecture generalizes a block-wise MBIR algorithm that uses momentum and majorizers for fast convergence without needing inner iterations~\cite{momen:19:il};
	it has been successfully applied to low-dose~\cite{momen:20:Ye} and sparse-view~\cite{momen:19:il} CT reconstruction.
	Different from the aforementioned INN methods that solve image reconstruction problems in low-dose or sparse-view CT, highly-undersampled MRI, and low-count PET, the proposed INN architecture is designed for image-domain material decomposition in DECT.
	The initial version of this work was presented in a conference~\cite{li2019bcdnet}, where we used an MBID cost function for the model-based image reconstruction module of BCD-Net, and demonstrated that BCD-Net significantly improved image quality over DECT-EP and DECT-ST.
	The initial BCD-Net work~\cite{li2019bcdnet} has a single-hidden layer or ``shallow" CNN (sCNN) architecture, where sCNN refiner has identical encoding-decoding architecture 
	individually for two different materials~(e.g., water and bone).
	The aforementioned INNs are trained in a supervised manner, whereas the recent study~\cite{N2N:21:quan} applied a self-supervised image denoising method to an INN.
	
	\subsection{Contributions}
	Image-domain material decomposition methods in DECT are susceptible to noise and artifacts (see Section~\ref{sec:background}). 
	Our aim is to obtain high-quality decomposed material images in DECT with improved image-domain material decomposition methods. To achieve the goal, the paper proposes an improved BCD-Net architecture.
	The proposed BCD-Net uses iteration-wise sCNN refiners, where they use \textit{1)} distinct encoding-decoding architecture, i.e., each filter in decoding convolution is distinct from that in encoding convolution, and \textit{2)} cross-material model that captures correlations between different material images.
	We refer to the previous BCD-Net in the earlier conference work~\cite{li2019bcdnet} as BCD-Net-sCNN-lc and the proposed BCD-Net in this work as BCD-Net-sCNN-hc, where lc and hc stand for low and high complexity, respectively.
	In addition, we study the proposed distinct cross-material CNN architecture with the patch-based perspective,
	empirically showing that learned filters of distinct cross-material CNN refiners at the last BCD-Net iteration approximately satisfy the tight-frame condition.
	The patch-based reformulation reveals that the proposed CNN architecture has the cross-material property, 
	and specializes to BCD-Net-sCNN-lc~\cite{li2019bcdnet} refiners.
	Our tight-frame studies imply that cross-material CNN refiners are useful for noise suppression and signal restoration.
	The quantitative and qualitative results with extended cardiac-torso (XCAT) phantom and clinical data show that the proposed BCD-Net-sCNN-hc architecture significantly improves the decomposition quality compared to the conventional MBID method, DECT-EP~\cite{xue:2017:statistical}, and the following recent image-domain decomposition methods, 
	a noniterative dCNN method and
	a MBID method, DECT-ST~\cite{li2018image}, that uses a learned regularizer in an unsupervised way, 
	and BCD-Net-sCNN-lc~\cite{li2019bcdnet}.

	\subsection{Organization}
	The rest of this paper is organized as follows. 
	Section~\ref{sec:methods} describes the proposed BCD-Net architecture for DECT image-domain MBID, 
	studies the distinct cross-material refining sCNN architecture with the patch-based reformulation and the tight-frame condition,
	and  provides training and testing algorithms for proposed BCD-Net architectures. 
	Section~\ref{sec:experiments} reports results of various decomposition methods on XCAT phantom and clinical data, along with comparisons and discussions.
	Finally, we make conclusions of this paper, and describe future work in Section~\ref{sec:conclusions}.
	
	\section{Methods}
	\label{sec:methods}
	This section proposes the BCD-Net-sCNN-hc architecture, studies properties of its refiners, introduces its variations, and describes its training and testing processes.

	\subsection{The Proposed BCD-Net Architecture}
	\label{sec:discro_refiner}
	Each iteration of BCD-Net for DECT material decomposition consists of an image refining module and an MBID module.
	See the architecture of the proposed BCD-Net in Figure~\ref{fig:dis_cro_framework}.
	Each image refining module of proposed BCD-Net has a sCNN architecture that consists of encoding convolution, nonlinear thresholding, and decoding convolution.
	The MBID cost function uses a weighted least-squares (WLS) data-fit term that models the material composition physics and noise statistics in the measurements, and a regularizer (or a prior term) that uses refined material images from an iteration-wise image refining module. 
	In DECT, decomposing high- and low-energy attenuation images into two material images (water and bone) is the most conventional setup~\cite{cle2009}, so the section studies the proposed INN method with this perspective.

	\subsubsection{Image Refining Module}
\begin{figure}[!t]
		\centering 
		\includegraphics[scale=0.4,trim=50 45 50 10,clip]{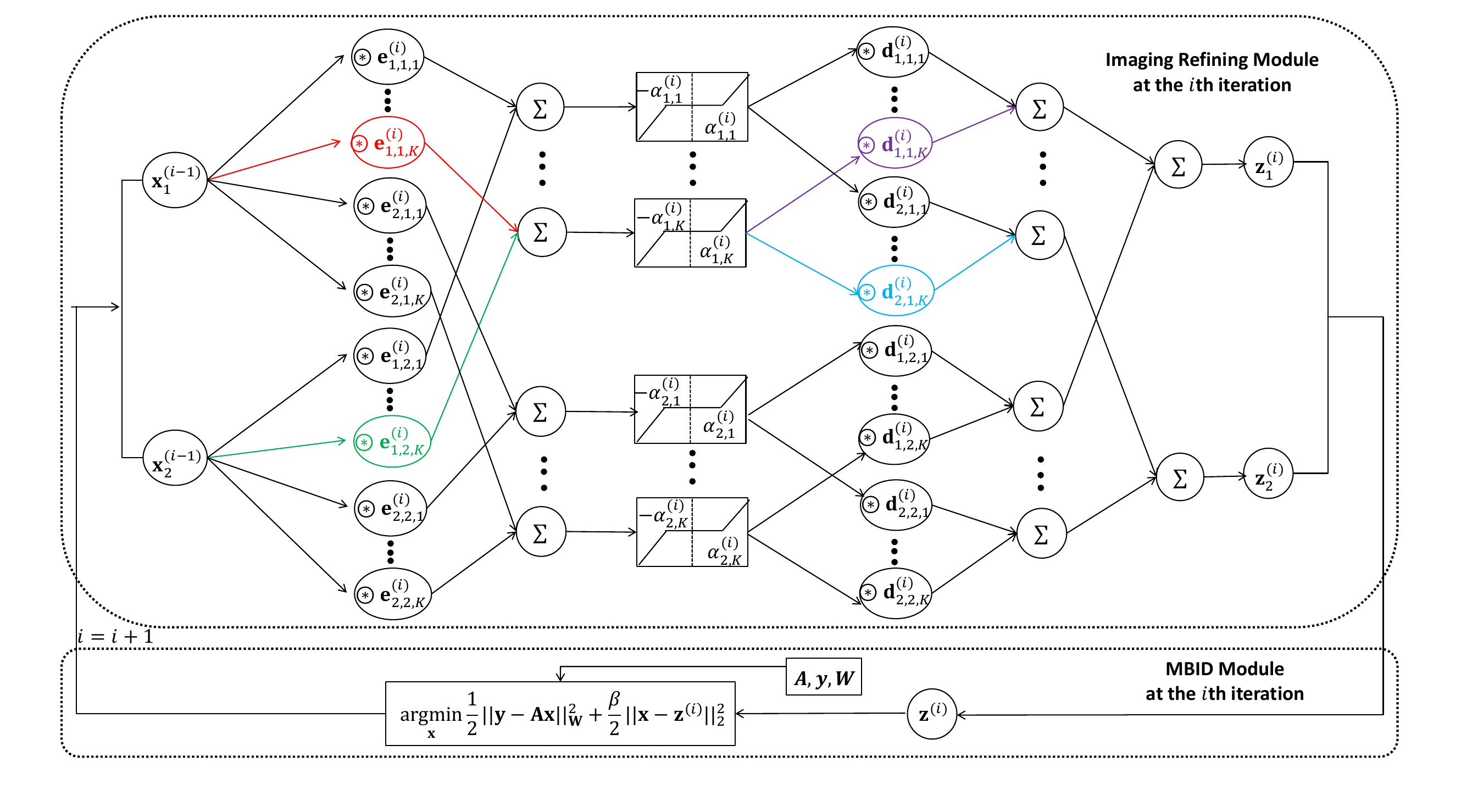}
		\caption{The proposed BCD-Net architecture at the $i$th iteration, for $i=1,\ldots,I_{\text{iter}}$.}
		\label{fig:dis_cro_framework}
		\vspace{-0.1in}
	\end{figure}

	The first box in Figure~\ref{fig:dis_cro_framework} shows the architecture of proposed iteration-wise distinct cross-material CNNs.
	The $i$th image refining module of BCD-Net takes $\{ \x_m^{(i-1)} \in \mathbb{R}^{N} : m = 1,2\}$, decomposed material images at the $(i-1)$th iteration, and outputs refined material images $\{ \z_m^{(i)} \in \mathbb{R}^{N} : m = 1,2 \}$, for $i = 1, \ldots, I_{\text{iter}}$, where $I_{\text{iter}}$ is the number of BCD-Net iterations. Here, $\{\x_1, \z_1\}$, and $\{\x_2, \z_2\}$ denote water and bone images, respectively.
	We use the following sCNN architecture for each image refining module:
	\begin{linenomath}
		\begin{equation}
		\begin{aligned}
		\label{eq:DisCro_Mapping}
		(\z_1^{(i)}, \z_2^{(i)}) = \mathcal{R}_{\Theta^{(i)}}\left( \x_1^{(i-1)}, \x_2^{(i-1)} \right)
		=  
		\hspace{-0.3em}\left[ \hspace{-0.4em}
		\begin{array}{c}
		\sum_{k=1}^{K}\sum_{n=1}^{2} \dd_{1,n,k}^{(i)} \circledast\TT_{\exp(\alpha_{n,k}^{(i)})} \left( \sum_{m=1}^{2} \ee_{n,m,k}^{(i)} \circledast \x_m^{(i-1)} \right) \\ 
		\sum_{k=1}^{K}\sum_{n=1}^{2} \dd_{2,n,k}^{(i)} \circledast \TT_{\exp(\alpha_{n,k}^{(i)})} \left( \sum_{m=1}^{2} \ee_{n,m,k}^{(i)} \circledast \x_m^{(i-1)} \right)
		\end{array}\hspace{-0.4em}
		\right]_,
		\end{aligned}
		\end{equation}
	\end{linenomath}
	where $\Theta^{(i)}$ denotes a set of parameters of image refining module at the $i$th iteration,
	i.e., $\Theta^{(i)} = \{ \dd_{m,n,k}^{(i)}, \ee_{n,m,k}^{(i)}, \alpha_{n,k}^{(i)}: k = 1,\ldots,K, m = 1,2, n = 1,2 \}$,
	$\dd_{m,n,k}^{(i)} \in \mathbb{R}^R$ and $\ee_{n,m,k}^{(i)} \in \mathbb{R}^R$ are the $k$th decoding and encoding filters from the $n$th group of the $m$th material at the $i$th iteration, respectively, $\exp(\alpha_{m,k}^{(i)})$ is the $k$th thresholding value for the $m$th material at the $i$th iteration,
	$K$ is the number of filters in each encoding and decoding structure for each material, and $R$ is the size of filters, $\forall m,n,k,i$.
	In (\ref{eq:DisCro_Mapping}), the element-wise soft thresholding operator $\TT_{\aaa}(\mathbf b) : \mathbb{R}^{N} \rightarrow \mathbb{R}^{N}$  is defined by
	\begin{linenomath}
		\begin{eqnarray}
		(\TT_{\aaa}(\mathbf b))_j:= 
		\begin{cases}
		b_j - a_j \cdot \mathrm{sign}(b_j) ,  &|b_j|  >  a_j \\
		0, &|b_j|   \leq a_j,
		\end{cases}
		\label{eq:threshold}
		\end{eqnarray}
	\end{linenomath}
		for $j=1,\ldots,N$.
	We use the exponential function to thresholding parameters $\{\alpha_{n,k}\}$ to avoid thresholding values being negative~\cite{bcdnet:19:il,momen:19:il}.
	We will train distinct cross-material CNNs at each iteration to maximize the refinement performance. 
	
	The proposed CNN in (\ref{eq:DisCro_Mapping}) and the first box in Figure~\ref{fig:dis_cro_framework} consists of an individual encoding-decoding architecture for each material image, and crossover architectures between different material images.
	We encode or decode each feature at a hidden layer by two groups of encoding or decoding filters.
	For example, in Figure~\ref{fig:dis_cro_framework}, input images $\x_1^{(i-1)}$ and $\x_2^{(i-1)}$ convolve with encoding filters $\ee_{1,1,K}^{(i)}$ and $\ee_{1,2,K}^{(i)}$, respectively (indicated by red and green), and then their thresholded sum gives encoded feature $\TT_{\exp(\alpha_{1,K}^{(i)})}(\ee_{1,1,K}^{(i)} * \x_1^{(i-1)}+\ee_{1,2,K}^{(i)} * \x_2^{(i-1)})$.
	To decode the feature, we convolve this feature with two decoding filters $\dd_{1,1,K}^{(i)}$ and $\dd_{2,1,K}^{(i)}$ (indicated by purple and blue). 
	One group of encoding or decoding filters is used to capture a feature of each material image individually, and the other group is used to capture correlations between different material images.
	When $n=m$, the filters in (\ref{eq:DisCro_Mapping})  form the individual encoding-decoding architecture that captures individual properties of the $m$th material,
	e.g., filters $\ee_{1,1,K}^{(i)}$ and $\dd_{1,1,K}^{(i)}$ (indicated by red and purple in Figure~\ref{fig:dis_cro_framework}),
	whereas when $n\neq m$, these comprise the crossover architecture that exchanges information between two material images,
	e.g., filters $\ee_{1,2,K}^{(i)}$ and $\dd_{2,1,K}^{(i)}$ (indicated by green and blue in Figure~\ref{fig:dis_cro_framework}).
	The crossover architecture is expected to be useful to remove noise and artifacts in material images.
	
	\subsubsection{MBID Module}
	\label{sec:mbid}
	The $i$th MBID module of BCD-Net in the second box of Figure~\ref{fig:dis_cro_framework} gives the decomposed material images, 
	$\x^{(i)} = [(\x_1^{(i)})^\top,(\x_2^{(i)})^\top]^\top$, by reducing their deviations from attenuation maps $\y = [(\y_H)^\top,(\y_L)^\top]^\top \in \mathbb{R}^{2N}$ and refined material images $\z^{(i)} = [(\z_1^{(i)})^\top,(\z_2^{(i)})^\top]^\top, \forall i$,
	where $\y_H \in \mathbb R^{N}$ and $\y_L \in \mathbb R^{N}$ are attenuation maps at high and low energy, respectively.
	In particular, we reduce the deviation of model-based decomposition $\x^{(i)}$ from attenuation maps $\y$, 
	using decomposition physics and noise statistics in $\y$.
	We formulate the MBID cost function by combining a WLS data-fit term and a regularizer using $\z^{(i)}$:
	\begin{linenomath}
		\begin{equation}
		\label{eq:MBID}
		\x^{(i)} = \underset{\x\in \mathbb{R}^{2N}}{\operatorname{argmin}} \frac{1}{2}\|\y - \A \x\|^2_{\W}  + \G(\x), \quad
		\G(\x) = \frac{\beta}{2} \|\x - \z^{(i)}\|^2_2.
		\tag{P0}
		\end{equation}
	\end{linenomath}
	The mass attenuation coefficient matrix $\A\in \mathbb R^{2N\times 2N}$  is a Kronecker product of $\A_0$ and identity matrix $\I_{N}$, i.e., $\A=\A_0\otimes\I_{N}$, and the matrix $\A_0\in \mathbb{R}^{2\times 2}$ is defined as~\cite{li2018image}:
	\vspace{-5pt}
	\begin{linenomath}
		\begin{equation}
		\A_0 :=  \Bigg[
		\begin{array}{cc}
		\varphi_{1H} & \varphi_{2H} \\
		\varphi_{1L} & \varphi_{2L} \\
		\end{array}
		\Bigg],
		\vspace{-3pt}
		\end{equation}
	\end{linenomath}
	in which $\varphi_{mH}$ and $\varphi_{mL}$ denote the mass attenuation coefficient of the $m$th material at high and low energy, respectively. In practice, these four values in matrix $\A_0$ can be calibrated in advance by $\varphi_{mH} = \mu_{mH}/ \rho_m$ and $\varphi_{mL} = \mu_{mL}/ \rho_m$, where $\rho_m$ denotes the density of the $m$th material (we use theoretical values 1 g/cm$^3$ for water and 1.92 g/cm$^3$ for bone in our experiments), and $\mu_{mH}$ and $\mu_{mL}$ denote the linear attenuation coefficient of the $m$th material at high and low energy, respectively. 
	To obtain $\mu_{mH}$ and $\mu_{mL}$, we manually select a uniform area in $\y_H$ and $\y_L$ (e.g., water region and bone region) respectively and compute the average pixel value in this area~\cite{niu2014iterative}. 
	The weight matrix $\W\in\mathbb{R}^{2N\times 2N}$ represented as $\W=\W_0\otimes \I_{N}$ is block-diagonal by assuming the noise in each attenuation image are independent and identically distributed (i.i.d.) over pixels~\cite{xue:2017:statistical}.
	This noise assumption is widely used in practice~\citep{xue:2017:statistical,xue:phantom,wu:dic,wu:19:dl}.
	Here, $\W_0$ is a $2\times 2$ diagonal weight matrix with diagonal elements being the inverse of noise variance at high and low energies.
	The regularization parameter $\beta>0$ controls the trade-off between noise and resolution in decompositions.
	
	Based on the structures of matrices $\A$ and $\W$ above, we can separate the \textup{$\x$}-update problem in (\ref{eq:MBID}) into $N$ subproblems. 
	Then we obtain the following practical closed-form solution of $\x$ at each pixel $j$:
	\vspace{-4pt}
	\begin{linenomath}
		\begin{equation}
		\label{eq:x_update}
		\x_j^{(i)} = (\A_0^\top \W_0 \A_0 + \beta \I_2)^{-1} (\A_0^\top \W_0 \y_j + \beta \z_j^{(i)}),
		\vspace{-4pt}
		\end{equation}
	\end{linenomath}
	where $\x_j^{(i)}=(x_{1,j}^{(i)},\,x_{2,j}^{(i)})^\top$ and $\z_j^{(i)}=(z_{1,j}^{(i)},\,z_{2,j}^{(i)})^\top$ denote the water and bone density values of decomposed material images $\x^{(i)}$ and refined material images $\z^{(i)}$ at the $j$th pixel, respectively, and $\y_j=(y_{H,j},\,y_{L,j})^\top$ denotes the high- and low-energy linear attenuation coefficients at the $j$th pixel,
	$j = 1,\ldots,N$.
	Due to small dimensions of matrices $\A_0^\top \W_0 \A_0$ and $\I_2$, the matrix inversion in (\ref{eq:x_update}) is efficient; the cost to compute $\{\x_j^{(i)}: \forall j\}$ scales as $O(N)$.
	Permuting $\{ \x_j^{(i)}: \forall j \}$ gives the decomposed material images $\x^{(i)}=(x_{1,1}^{(i)},\dots,x_{1,N}^{(i)},x_{2,1}^{(i)},\dots,x_{2,N}^{(i)})^\top$.
	
	
	\subsection{Properties of the Proposed CNN Refiner}
	\label{sec:Interpretation}
	This section studies some properties of the proposed CNN~(\ref{eq:DisCro_Mapping}) with the patch perspective.
	We rewrite~(\ref{eq:DisCro_Mapping}) with the patch perspective as follows (we omit the iteration superscript indices $(i)$ for simplicity):
	\begin{linenomath}
		\begin{equation}
		\mathcal{R}_{\Theta}(\x) \textup{ in (\ref{eq:DisCro_Mapping})}
		= \frac{1}{R}\sum_{j=1}^{N} \bar{\P}_j^\top \D \TT_{\exp(\alp)} (\E\bar{\P}_j \x),
		\label{eq:conv2patch}
		\end{equation}
	\end{linenomath}
	where,  $\bar{\P}_j = \P_j \oplus \P_j$,  $\P_j \in \mathbb{R}^{R\times N}$ is the patch extraction operator for the $j$th pixel, $j=1,\dots,N$, $\oplus$ denotes the matrix direct sum,
	$\D \in \mathbb{R}^{2R\times 2K}$ and $\E \in \mathbb{R}^{2K\times 2R}$ are decoding and encoding filter matrices defined by:
	\begin{linenomath}
		\begin{equation}
		\centering
		\D:=\left[ \begin{array}{cc}
		\D_{1,1} & \D_{1,2} \\
		\D_{2,1} & \D_{2,2}
		\end{array} \right] \hspace{0.05in}
		\textup{and}  \hspace{0.07in}
		\E:=\left[ \begin{array}{cc}
		\E_{1,1} & \E_{1,2} \\
		\E_{2,1} & \E_{2,2}
		\end{array} \right],
		\label{eq:D-and-E}
		\end{equation} 
	\end{linenomath}
	where $\D_{m,n}$ and $\E_{n,m}$ are formed by grouping filters $\{\dd_{m,n,k}\}$ and $\{\ee_{n,m,k}\}$, respectively, i.e., 
	\begin{linenomath}
		\begin{equation*}
		\begin{aligned}
		& \D_{m,n}:=\left[\dd_{m,n,1},\,\dd_{m,n,2}, \dots, \dd_{m,n,K}\right], \\
		& \E_{n,m}:=\left[\ee_{n,m,1},\,\,\ee_{n,m,2},\, \dots, \ee_{n,m,K}\right]^\top, \quad m,n=1,2,
		\end{aligned}
		\end{equation*}
	\end{linenomath}
	and $\alp=[\alpha_{1,1},\dots,\alpha_{1,K},\alpha_{2,1},\dots,\alpha_{2,K}]^\top \in \mathbb{R}^{2K}$ is a vector containing $2K$ thresholding parameters.
	We derived (\ref{eq:conv2patch}) using the convolution-to-patch reformulation technique~\cite{momen:19:il}; see Proposition~\ref{prop:refiner} for more details.

	Both of encoding and decoding filter matrices, $\E$ and $\D$,  are composed of four smaller block matrices.
	The refiner of BCD-Net-sCNN-lc~\cite{li2019bcdnet} uses only block matrices $\E_{1,1}$ and $\E_{1,1}^\top$ as encoding and decoding filters, respectively, for water images,  
	and $\E_{2,2}$ and $\E_{2,2}^\top$ as the encoding and decoding filters, respectively, for bone images. 
	Different from this, the proposed refiner of BCD-Net-sCNN-hc 
	not only uses \emph{distinct} encoding-decoding filters, but also 
	additionally uses off-diagonal block matrices $\{\D_{1,2}, \D_{2,1}, \E_{1,2}, \E_{2,1}\}$ to exploit correlations between the different material images. 
	The crossover architecture captured via $\{\D_{1,2}, \D_{2,1}, \E_{1,2}, \E_{2,1}\}$
	models shared structures between water and bone images at the same spatial locations.
	When trained with some image denoising loss, the crossover architecture with thresholding operator (\ref{eq:threshold}) in BCD-Net-sCNN-hc
	is expected to better refine material images by exchanging shared noisy features between them, 
	compared to the individual encoding-decoding case in BCD-Net-sCNN-lc.

\begin{figure}[!t]
		\centering
		\includegraphics[scale=0.45]{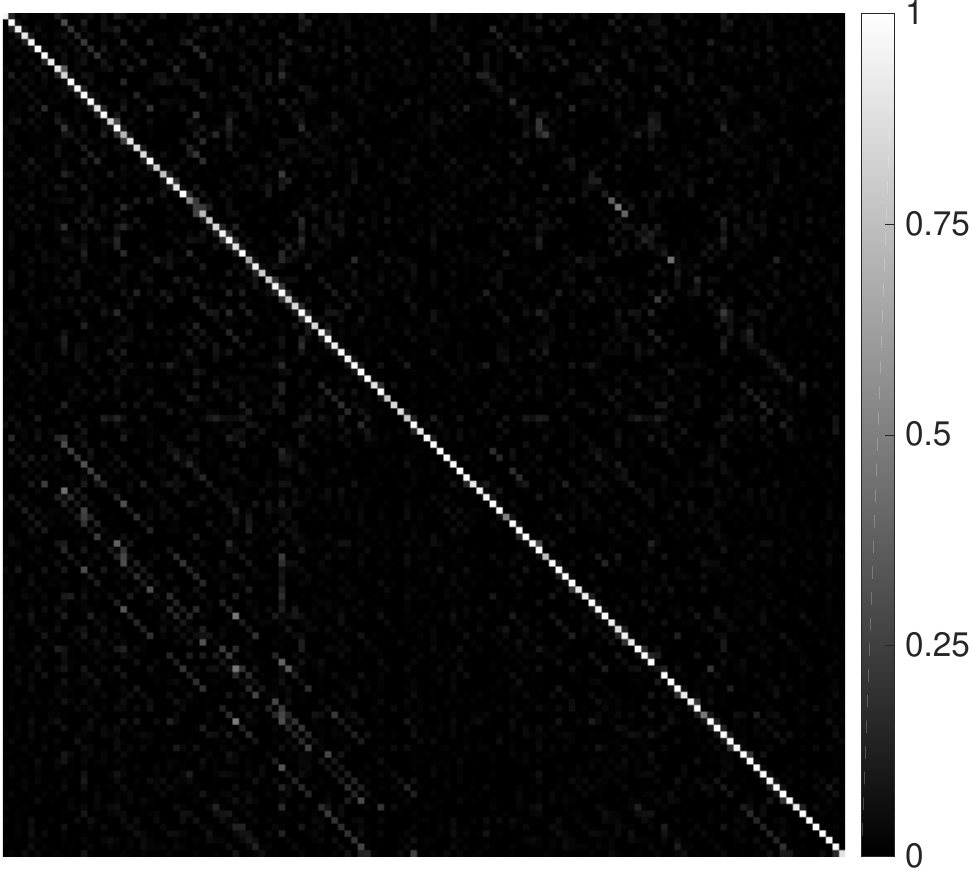} 
		\vspace{-6pt}
		\caption{$\D^{(100)} \E^{(100)}$ of BCD-Net-sCNN-hc.}
		\label{fig:tf}
	\end{figure}

	We study the tight-frame property~\cite{waldron2018introduction} of the proposed cross-material CNN refiners, 
	since learned filters satisfying the tight-frame condition are useful to compact energy of input image and remove unwanted noise and artifacts via thresholding~\cite{cai2014data,CAOL:19:il}.
	The tight-frame condition for (\ref{eq:conv2patch}) is given by 
	\begin{linenomath}
		\begin{equation}
		\centering
		\D\E=\I_{2R}.
		\label{eq:tf-cond}
		\end{equation}
	\end{linenomath}	
	This is implied as follows.
	Using the patch-perspective reformulation~(\ref{eq:conv2patch}), convolutional encoding in~(\ref{eq:DisCro_Mapping}) can be rewritten as follows:
		$\sqrt{1/R}	[  (  \E \bar{\P}_1 )^\top, \dots,  \left(  \E \bar{\P}_N\right)^\top ]^\top \x.$
	The tight-frame condition for a refiner that uses this as both encoder and decoder, i.e.,~(\ref{eq:IdInd_Mapping}) in Section~\ref{sec:variations}, is given as follows~\cite{CAOL:19:il,cai2014data}:
	$	\|\x\|^2 =  \x^\top \sum_{j=1}^{N} \bar{\P}_j^\top \E^\top \E \bar{\P}_j \x / R, \quad \forall \,\x.$
		\label{eq:tf}
	This condition is identical to $\E^\top \E = \I_{2R}$ considering that $\sum_{j=1}^{N}\bar{\P}_j^\top \bar{\P}_j=R\I_{2N}$ with the periodic boundary condition and sliding parameter 1.
	If a decoding filter matrix is different from an encoding filter matrix, e.g., (\ref{eq:DisCro_Mapping}), then the tight-frame condition can become (\ref{eq:tf-cond}).
	In Figure~\ref{fig:tf}, we empirically observed for DECT material decomposition that sCNN-hc refiners of BCD-Net at the last iteration approximately satisfy the tight-frame condition. 
	
\begin{figure}[!t]
		\centering
		\includegraphics[scale=1.33]{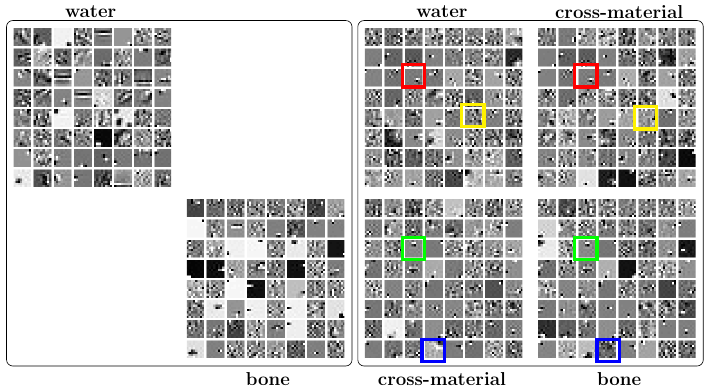}
		\caption{Left and right are learned filters of BCD-Net-sCNN-lc and BCD-Net-sCNN-hc at the last iteration that uses identical encoding-decoding architecture (i.e., $\D=\E^\top$), respectively.
			Top-left, top-right, bottom-left, and bottom-right correspond to $\E_{1,1}$, $\E_{1,2}$, $\E_{2,1}$, and $\E_{2,2}$, respectively. 
			Four pairs of filters (indicated by four different colors) are selected as examples to show similar or different structures between off-diagonal and diagonal blcok matrices;
			filters indicated by red or green boxes show similar structures, while blue or yellow boxes show different structures.
		}	
		\label{fig:filters-comp}	
	\end{figure}

	Figure~\ref{fig:filters-comp} shows learned filters of BCD-Net-sCNN-lc and  BCD-Net-sCNN-hc refiners that use the identical encoding-decoding architecture, i.e., $\D=\E^\top$ in (\ref{eq:conv2patch}), where we display them with four groups, $\E_{1,1}$, $\E_{1,2}$, $\E_{2,1}$, and $\E_{2,2}$ in~(\ref{eq:D-and-E}). 
	Filters in diagonal block matrices on the left in Figure~\ref{fig:filters-comp} include both (short) first-order finite differences and elongated features.
	In addition, $\E_{1,1}$ includes more elongated structures than $\E_{2,2}$, while $\E_{2,2}$ includes more first-order finite difference like kernels than $\E_{1,1}$ (there are 16 and 23 first-order finite difference like structures in $\E_{1,1}$ and $\E_{2,2}$, respectively).
	This is potentially because water image includes diverse low-contrast edge features from different soft-tissues, while bone image includes relatively simple high-contrast edge features from bone and air.
	Many structured kernels in $\E_{1,1}, \E_{1,2}, \E_{2,1}$, and $\E_{2,2}$, on the right in Figure 3 are like first-order finite difference: specifically, $\E_{1,1}$, $\E_{1,2}$, $\E_{2,1}$, and $\E_{2,2}$ have about 10, 17, 17, and 24 first-order finite difference like kernels.
	Interestingly, the number of first-order finite difference like kernels of $\E_{1,2}$ and $\E_{2,1}$ is  intermediate between those of $\E_{1,1}$ and $\E_{2,2}$. 
	This might imply using the conjecture above that cross-materials have less and more diverse edge features than water image and bone image, respectively.
	What is more, we observed some filters in $\E_{1,2}$ capture similar features as those in $\E_{1,1}$, e.g., filters indicated by red boxes,  
	while some filters in $\E_{1,2}$ capture different features from those in $\E_{1,1}$, e.g., filters indicated by yellow boxes.
	We also observed similar behavior between $\E_{2,1}$ and $\E_{2,2}$.
	
	\subsection{Variations of (\ref{eq:DisCro_Mapping})}
	\label{sec:variations}
	We specialize~(\ref{eq:DisCro_Mapping})  to have simpler components.
	BCD-Net-sCNN-lc is a simpler convolutional encoding-decoding architecture proposed in our recent conference work~\cite{li2019bcdnet};
	it uses following CNN refiner that has identical encoding-decoding architecture independently for two different material images:
	\begin{linenomath}
		\begin{equation}
		\z^{(i)}_m=\mathcal{R}_{\Theta_m^{(i)}}(\x^{(i-1)}_m) = \sum_{k=1}^{K} {\bar{\ee}}^{(i)}_{m,m,k} \circledast \TT_{\exp{(\alpha^{(i)}_{m,k})}} \left({\ee_{m,m,k}^{(i)}} \circledast \x^{(i-1)}_m \right), \quad m=1,2,
		\label{eq:IdInd_Mapping}
		\end{equation}
	\end{linenomath}
	where $(\bar{\cdot})$ rotates a filter (e.g., it rotates 2D filters by 180$^\circ$).
	(\ref{eq:DisCro_Mapping}) specializes to (\ref{eq:IdInd_Mapping})  by setting $\dd_{m,n,k}^{(i)}$ as $\bar{\ee}_{n,m,k}^{(i)}$, and $\ee_{n,m,k}^{(i)}=\dd_{m,n,k}^{(i)}=\mathbf{0}$ for $m\neq n$.	
	One can also use dCNNs instead of the sCNN refiners in (\ref{eq:DisCro_Mapping}) and (\ref{eq:IdInd_Mapping}). 
	We refer to this method as BCD-Net-dCNN.
	We investigate the  performance of BCD-Net-dCNN (that replaces the refining module in (\ref{eq:DisCro_Mapping}) and (\ref{eq:IdInd_Mapping}) with a dCNN);
	see Section~\ref{sec:parameters} later for details of BCD-Net-dCNN.
	
	\subsection{Training BCD-Net-sCNNs}
	\label{sec:Traing}

	The training process at the $i$th iteration requires $L$ input-output image pairs.
	Input labels are decomposed material images via MBID module, $\{\x_{l,m}^{(i-1)}: l=1,\cdots,L\}$,
	and output labels are high-quality reference material images, $\{\x_{l,m}: l=1,\cdots,L\}$. 
	We use the patch-based training loss of  $(1/L)\sum_{l=1}^{L}\|\x_l-\mathcal{R}_\Theta(\x_l^{(i-1)})\|^2_2$, where we derived their bound relation in Proposition~\ref{prop:loss} using the convolution-to-patch loss reformulation techniques in a recent work~\cite{momen:19:il}.  
	Patch-based training first extracts reference and noisy material patches from $\{\x_{l,m}: l=1,\cdots,L\}$ and $\{\x_{l,m}^{(i-1)}: l=1,\cdots,L\}$ and constructs reference and noisy material data matrices $\widetilde{\X}_m \in \mathbb{R}^{R\times P}$ and $\widetilde{\X}_m^{(i-1)}\in\mathbb{R}^{R\times P}$, respectively, where $P=LN$.
	(For $\{ \x_{l,m}^{(0)}: \forall l, m \}$, we used rough estimates of decomposed images obtained via the direct matrix inversion method (see Section~\ref{sec:direct-matrix-inversion}).)
	Then we construct paired multi-material data matrices $\widetilde{\X}\in\mathbb{R}^{2R\times P}$ and $\widetilde{\X}^{(i-1)}\in\mathbb{R}^{2R\times P}$, where each column is formed by stacking vectorized two-dimensional (2D) patches extracted from the same spatial location in different material images.
	i.e., $\widetilde{\X}=[\widetilde{\X}_1^\top, \widetilde{\X}_2^\top]^\top$ and $\widetilde{\X}^{(i-1)}=[(\widetilde{\X}_1^{(i-1)})^\top, (\widetilde{\X}_2^{(i-1)})^\top]^\top$.
	
	\begin{algorithm}[!t]
		\caption{Training BCD-Net-sCNN-hc}  
		\begin{algorithmic}
			\REQUIRE \hspace{-6pt} $\{ \x_{l,m}, \x_{l,m}^{(0)}, \y_l, \A_l, \W_l: l = 1,\ldots,L, m = 1,2 \}, \beta>0$, $I_\mathrm{iter}>0$
			\bindent
			\FOR{$i = 1,2,\cdots, I_\mathrm{iter} $}
			\STATE{ Train $\Theta^{(i)}$ via (\ref{eq:DisCro_Loss}) using $\{\x_{l,m}, \x_{l,m}^{(i-1)} : \forall l,m\}$  }
			\FOR {$l=1,\dots,L$}
			\STATE \textbf{Refining:} $(\z_{l,1}^{(i)},\,\z_{l,2}^{(i)}) = \mathcal{R}_{\Theta^{(i)}} (\x_{l,1}^{(i-1)},\,\x_{l,2}^{(i-1)})$  in (\ref{eq:DisCro_Mapping}).
			\STATE \textbf{MBID:} Obtain $\{ \x_{l,m}^{(i)}: \forall l,m \}$ by solving (\ref{eq:MBID}) with (\ref{eq:x_update}).	
			\ENDFOR 					
			\ENDFOR 
			\eindent
		\end{algorithmic}
		\label{alg:1}
	\end{algorithm}
	
	The training loss of BCD-Net-sCNN-hc at the $i$th iteration is
	\begin{linenomath}
		\begin{equation}
		\label{eq:DisCro_Loss}
		\mathcal{L}(\D,\,\E,\,\alp) := \frac{1}{P}  \|  \widetilde{\X} -  \D  \TT_{\exp(\alp)}  (\E\widetilde{\X}^{(i-1)}) \|_{\mathrm{F}}^2,
		\tag{P1}
		\end{equation}
	\end{linenomath}
	where $\| \cdot \|_\mathrm{F}$ denotes the Frobenius norm of a matrix. The subgradients of $\mathcal{L}(\D,\,\E,\,\alp)$ with respect to $\D$, $\E$, and $\alp$   for each mini-batch selection are as follows:
	\begin{linenomath}
		\begin{equation}
		\label{eq:grad_D}
		\frac{\partial \mathcal{L}(\D,\,\E,\,\alp)}{\partial \D}= -\frac{2}{B}\left(\X-\D\Z ^{(i-1)}\right){\Z^{(i-1)}}^\top
		\end{equation}
	\end{linenomath}
	\begin{linenomath}
		\begin{equation}
		\label{eq:grad_W}
		\frac{\partial \mathcal{L}(\D,\,\E,\,\alp)}{\partial \E} = -\frac{2}{B} \D^\top\left(\X - \D\Z^{(i-1)}\right) \odot  \mathbbm{1}_{|\E\X^{(i-1)}|>\exp{(\alp\mathbf{1'})}} \cdot {\X^{(i-1)}}^\top
		\end{equation}
	\end{linenomath}
	\begin{linenomath}
		\begin{equation}
		\frac{\partial \mathcal{L}(\D,\,\E,\,\alp)}{\partial \alp} = \frac{2}{B}\left\{\D^\top   \left(\X-\D \Z^{(i-1)} \right)   \odot  \right.\left. \exp(\alp \mathbf 1') \odot  \textup{sign}\left(\Z^{(i-1)}\right) \right\}\mathbf{1},
		\label{eq:grad_gam}
		\end{equation}
	\end{linenomath}
	where 
	$\X,\, \X^{(i-1)} \in \mathbb{R}^{2R\times B}$ are mini-batch in which columns are randomly selected from $\widetilde{\X}$ and $\widetilde{\X}^{(i-1)}$, respectively,  
	$\Z^{(i-1)}=\TT_{\exp(\alpr)}  (\E\X^{(i-1)})$, and $B$ is the mini-batch size. 
	Here, $\mathbf{1}\in \mathbb{R}^{B\times 1}$ denotes a column vector of ones, $\mathbbm{1}_{(\cdot)}$ is the indicator function (value 0 when condition is violated and 1 otherwise), and $\odot$ is the element-wise multiplication. 
	The derivation details of \labelcref{eq:grad_D,eq:grad_W,eq:grad_gam} are in Section~\ref{sec:subdiff}. 
	Once we obtain the learned filters and thresholding values, we apply them to refine material images. 
	These refined images are then fed into the MBID module.
	Algorithm \ref{alg:1} shows the training process of BCD-Net-sCNN-hc.

	Training BCD-Net-sCNN-lc only involves submatrices $\E_{1,1}^{(i)}$ and $\E_{2,2}^{(i)}$,  i.e.,  $\E_{1,2}^{(i)}=\E_{2,1}^{(i)}=\D_{1,2}^{(i)}=\D_{2,1}^{(i)}=\mathbf{0}$, $\D_{1,1}^{(i)}={\E_{1,1}^{(i)}}^\top$, and $\D_{2,2}^{(i)}={\E_{2,2}^{(i)}}^\top$ in (\ref{eq:DisCro_Loss}), and we train it using image pair $(\widetilde{\X}_m,\,\widetilde{\X}_m^{(i-1)})$, $\forall m,i$.
	See subgradients for training BCD-Net-sCNN-lc in our earlier conference work~\cite{li2019bcdnet}.

	\subsection{Testing Trained BCD-Nets}
	 At the $i$th iteration of BCD-Net-sCNN-hc, we apply learned filters and thresholding parameters $\Theta^{(i)}$ to noisy material images $\{ \x_m^{(i-1)}: m=1,2\}$ to obtain refined material images $\z^{(i)} = \mathcal{R}_{\Theta^{(i)}} (\x_{1}^{(i-1)},\,\x_{2}^{(i-1)})$,
	 where the definition of $\z^{(i)}$ is given in Section~\ref{sec:mbid}.
	 We then feed these refined images into the MBID module to obtain decomposed material images $\{ \x_m^{(i)}: m=1,2\}$.
	After some fixed iterations (where $I_\mathrm{iter}$ is chosen in training), BCD-Net-sCNN-hc gives the final decomposed images $\{\x_m^{(I_\mathrm{iter})} : m = 1,2\}$.
	Algorithm~\ref{alg:2} summarizes the test process of learned BCD-Net-sCNN-hc.
	The test process of BCD-Net-sCNN-lc and BCD-Net-dCNN are similar to that of BCD-Net-sCNN-hc.
	

	\section{Results and Discussions}
	\label{sec:experiments}
	This section describes experimental setup and reports comparison results with XCAT phantom~\cite{Segars2008Realistic} and clinical DECT head data.
	We compared the performances of three BCD-Net methods  (BCD-Net-sCNN-lc~\cite{li2019bcdnet}, BCD-Net-sCNN-hc, and BCD-Net-dCNN),
	the conventional direct matrix inversion method,  MBID methods using data-driven and conventional non-data-driven regularizers, DECT-ST~\cite{li2018image} and DECT-EP~\cite{xue:2017:statistical},
	and a (noniterative) dCNN method.
	
	\vspace{-6pt}
	\begin{algorithm}[!t]	
	\caption{Testing Trained BCD-Net-sCNN-hc}
	\renewcommand{\algorithmicrequire}{\textbf{Input:}} 
	\renewcommand{\algorithmicensure}{\textbf{Output:}}
		\begin{algorithmic}
			\REQUIRE \hspace{-6pt} $\{  \x_{m}^{(0)} : m = 1,2 \}, \y, \A,\W, \{ \Theta^{(i)}: i=1,\dots,I_\mathrm{iter} \}, \beta>0$
			\ENSURE \hspace{-6pt} $\{  \x_{m}^{(I_\mathrm{iter})}: m = 1,2 \}$
			\bindent
			\FOR{$i = 1,2,\cdots, I_\mathrm{iter}$}
			\STATE \textbf{Refining:} $(\z_{1}^{(i)},\,\z_{2}^{(i)}) = \mathcal{R}_{\Theta^{(i)}} (\x_{1}^{(i-1)},\,\x_{2}^{(i-1)})$  in (\ref{eq:DisCro_Mapping}).
			\STATE \textbf{MBID:} Obtain $\{ \x_{m}^{(i)}: m=1,2 \}$ by solving (\ref{eq:MBID}) with (\ref{eq:x_update}).	
			\ENDFOR 					
			\eindent
		\end{algorithmic}
		\label{alg:2}
	\end{algorithm}

	\subsection{Methods for Comparisons}
	\label{subsec:MethCom}
	This section describes methods compared with the proposed BCD-Net methods.
	We will describe their parameters in the next section.
	\subsubsection{Direct Matrix Inversion }
	\label{sec:direct-matrix-inversion}
	This conventional method solves (\ref{eq:MBID}) with $\G(\x)=0$ by matrix inversion, i.e., $\A^{-1}\y$.
	We use direct matrix inversion results as initial material decomposition to DECT-EP and BCD-Nets, i.e., $\{\x^{(0)} = \A^{-1} \y \}$,
	and noisy input material images to dCNN denoiser.

	\subsubsection{DECT-EP }
	This conventional method solves  (\ref{eq:MBID}) with a material-wise edge-preserving regularizer that is defined as $\G_{\textup{EP}}(\x)=\sum_{m=1}^{2} \beta_m \G_m(\x_m)$, where the $m$th material regularizer is 
	$\G_m(\x_m)=\sum_{j=1}^{N} \sum_{k\in S} \psi_m(x_{m,j}-x_{m,k})$, and $S$ is a list of indices that correspond to neighboring pixels 
	of a pixel $x_{m,j}$ with $|S|=R_{\textup{EP}}$, $\forall m,j$, where $R_{\textup{EP}}$ denotes the number of neighbors for each pixel.
	Here, the potential function is $\psi_m(t)\triangleq \frac{\delta^2_m}{3}(\sqrt{1+3(t/\delta_m)^2}-1)$ with the $m$th material EP parameter, $\delta_m$. 
	We chose $\beta_m$ and $\delta_m$ for different materials separately to achieve the desired boundary sharpness and strength of smoothness.
	
	\subsubsection{DECT-ST}
	This data-driven method solves (\ref{eq:MBID}) with a regularizer that uses two square material-wise sparsifying transforms trained in an unsupervised way.
	The regularizer $\G_{\textup{ST}}(\x)$ is defined as
	\vspace{-0.05in}
	\begin{linenomath}
		\begin{equation*}
		\label{Eq:DECT_ST} \hspace{-0.07in}
		\G_{\textup{ST}}(\x)\hspace{-0.02in} \triangleq \hspace{-0.04in} \min \limits_{ \{ \z_{m,j} \} } \hspace{-0.03in} \sum_{m=1}^2 \hspace{-0.03in}  \sum_{j=1}^{N} \hspace{-0.03in} \beta_m \hspace{-0.03in}  \left\{ \left\| \omg_m  \P_{m,j}  \x \!-\! \z_{m,j} \right\|_2^2 \hspace{-0.025in}  + \hspace{-0.025in}  \gamma_m^2 \hspace{-0.02in}  \left\|  \z_{m,j}\right\|_0 \right\}\hspace{-0.03in},
		\vspace{-0.05in}
		\end{equation*}
	\end{linenomath}
	where $\omg_1\in\mathbb{R}^{R_\textup{ST}\times R_\textup{ST}}$ and $\omg_2\in\mathbb{R}^{R_\textup{ST}\times R_\textup{ST}}$ are pre-learned transforms for water and bone, respectively, $\P_{m,j}\x$ and $\z_{m,j}$ denote the $j$th patch of the $m$th material image and corresponding sparse vector, respectively, and $R_\textup{ST}$ is the number of pixels in each patch. 
	
	\subsubsection{dCNN denoiser}
	\label{sec:dcnn-denoiser}
	The (noniterative) image denoising dCNN method uses two input and output channels; specifically,  it takes noisy water and bone images and provides denoised water and bone images. 
	The architecture that maps from noisy material images to true material images corresponds to the second CNN architecture of the cascaded dCNN~\cite{YuCNN}, and that uses two input and two output channels corresponds to the setup of a modified U-Net method~\cite{clark2018multi}.
	
	\subsection{Experimental Setup}
	\label{subsec:ExSetup}
	
	\subsubsection{Imaging setup for XCAT phantom experiments}
	We used $1024\times1024$ material images with pixel size $0.49\times 0.49~$mm$^2$ of the XCAT phantom in our imaging simulation.   
	We generated noisy (Poisson noise) sinograms of size $888$ (radial samples) $ \times\, 984$ (angular views) using GE LightSpeed X-ray CT fan-beam system geometry corresponding to a poly-energetic
	source at 80~kVp and 140~kVp with $1.86\times 10^5$ and $1\times 10^6$ incident photons per ray, respectively. 
	We used FBP method to reconstruct 2D high- and low-energy attenuation images of size $512\times 512$ with a coarser pixel size $0.98\times0.98$~mm$^2$ to avoid an inverse crime.
	Figure~\ref{fig:atten_xcat} displays the attenuation images for a test slice.

	\subsubsection{Data construction}
	We separated each $1024\times 1024$ slice of the original XCAT phantom into water and bone images according to the table of linear attenuation coefficients for organs provided for the XCAT phantom.
	We manually grouped fat, muscle, water, and blood into the water density images, and rib bone and spine bone into bone density images.
	We then downsampled these material density images to size $512\times512$ by linear averaging to generate ground truths of the decomposed material images. 
	We chose 13 slices from the XCAT phantom, among which $L=10$ slices were used for training the proposed BCD-Net-sCNNs, and remaining 3 slices were used for testing.
	Testing phantom images are sufficiently different from training phantom images; 
	specifically, they are at a minimum $\approx 1.5$~cm away, i.e., 25 slices.
	For dCNN, we used $L=20$ slices of XCAT phantom that includes the 10 slices chosen for  training the proposed BCD-Net-sCNNs.
	In general, dCNNs need many training samples, so we used more image pairs to train dCNN compared to BCD-Net-sCNN-lc and BCD-Net-sCNN-hc.
	
	
	In addition, using the clinical data, we evaluated the proposed methods and compared them to the methods in Section~\ref{subsec:MethCom}.
	The clinical data experiments decomposed a mixture into two constituent materials, water and bone, in each pixel.
	The patient head data was obtained by Siemens SOMATOM Definition flash CT scanner using dual-energy CT imaging protocols. 
	The protocols of this head data acquisition are listed in Table~\ref{Tab:clinical}.
	For dual-energy data acquisition, the dual-energy source were set at 140~kVp and 80~kVp. 
	Figure~\ref{fig:decom_head} shows attenuation images of head data.
	FBP method was used to reconstruct these attenuation images.

\subsubsection{Methods setup and parameters}
\label{sec:parameters}
We first obtained the low-quality material images from high- and low-energy attenuation images using direct matrix inversion method, and used these results to initialize DECT-EP method.
We used the 8-neighborhood system, $R_{\textup{EP}}=8$.
To ensure convergence, we ran DECT-EP with 500 iterations. 
For XCAT phantom, we set \{$\beta_m,\,\delta_m:\, m=1,2$\} as \{$2^{8},\,0.01$\} and \{$2^{8.5},\,0.02$\} for water and bone, respectively; for patient head data, we set them as  $\{2^{10.5},\,0.008\}$ and $\{2^{11},\,0.015\}$ for water and bone, respectively.  

We pre-learned two sparsifying transform matrices of size $R_{\textup{ST}}^2 = 64^2$ with ten slices (same slices as used in training BCD-Net-sCNNs) of true water and bone images of the XCAT phantom, using the suggested algorithm and parameter set (including number of iterations, regularization parameters, transform initialization, etc.) in the original paper~\cite{li2018image}.
We initialized DECT-ST using decomposed images obtained by DECT-EP method. 
We tuned the parameters \{$\beta_1,\,\beta_2,\,\gamma_1,\,\gamma_2$\} and set them as $\{50,\,70,\,0.03,\,0.04\}$ for XCAT phantom, and $\{150,\,200,\,0.012,\,0.024\}$ for patient head data.

For the denoising dCNN architecture, we set the number of layers and number of features in hidden layers as 4 and 64, respectively.
We did not use batch normalization and bias because the pixel values of different training/testing images are of the same scale.
We learned the dCNN denoiser $\mathcal{R}$ with the standard loss in image denoising, $\mathcal{L}({\mathcal{R}})=\frac{1}{L} \sum_{l=1}^{L}\|\x_l-\mathcal{R}(\x_l^{(0)})\|_2^2$, 
with Adam using 200 epochs and batch size 1.
We observed with the clinical data that selected dCNN architecture gives better decomposed image quality,
compared to its variants with 8 layers and/or the different mode that maps high- and low-energy attenuation images to two material images
(this mode corresponds to a series of papers~\cite{Xu:2018:ImgCNN,Niu:18:butterfly,clark2018multi}). 	

\begin{figure}[!t]
		\centering
		\includegraphics[scale=1]{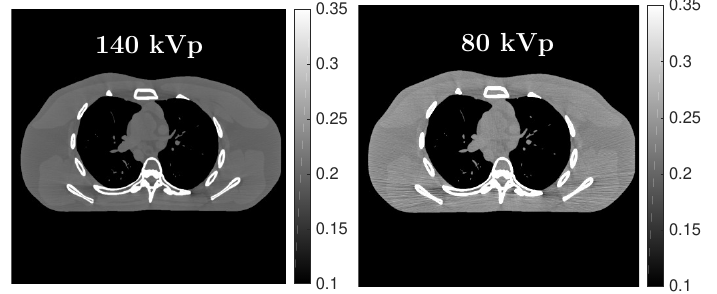}
		\vspace{-3pt}
		\caption{The attenuation images (zoomed-in) for a test slice at high and low energies, respectively.}
		\label{fig:atten_xcat}
		\vspace{-6pt}
	\end{figure}

We trained a 100-iteration BCD-Net-sCNN-hc and a 100-iteration BCD-Net-sCNN-lc with image refining CNN architectures in (\ref{eq:DisCro_Mapping}) and (\ref{eq:IdInd_Mapping}), respectively.
For BCD-Net-sCNN-hc, we trained cross-material CNN refiners in~(\ref{eq:DisCro_Mapping}) with about $1\times 10^6$ paired stacked multi-material patches.
We trained $8K=512$ filters of size $R=8\times 8$ at each iteration.
For BCD-Net-sCNN-lc, we trained convolutional refiners in~(\ref{eq:IdInd_Mapping}) for each material with about $1\times 10^6$ paired patches. 
We trained $K=64$ filters of size $R=8\times 8$ for each material at each iteration.
We initialized all filters with values randomly generated from a Gaussian distribution with a zero mean and standard deviation of 0.1.
We found in training that thresholding value initialization is important to ensure stable performances.
For BCD-Net-sCNN-lc, we set initial thresholding parameters before applying the exponential function as $\log(0.88)$ and $\log(0.8)$ for water and bone, respectively; for BCD-Net-sCNN-hc, we set them as $\log(0.88)$.
The regularization parameter $\beta$ balances data-fit term and the prior estimate from image refining module. 
To achieve the best image quality and decomposition accuracy,  
we set  $\beta$ as $600$ and $6400$ for BCD-Net-sCNN-lc and BCD-Net-sCNN-hc, respectively (note that different BCD-Net architectures have different refining performance). 
We train NNs of BCD-Net-sCNN-hc and BCD-Net-sCNN-lc with Adam~\cite{Kingma2014Adam} using the default hyper-parameters and tuned learning rate of $3\times 10^{-4}$.
We applied the learning rate schedule that decreases learning rates by a ratio of 90\% every five epochs. 
We set batch size and number of epochs as $B=10000$ and 50, respectively. 
For patient head data, we used the learned filters and thresholding values with XCAT phantom.
The attenuation maps of XCAT phantom and clinical head data were generated by different energy spectrum and dose, and the clinical head data is much more complex than the XCAT phantom (see Figures~\ref{fig:atten_xcat} and \ref{fig:decom_head}).
We thus set different regularization parameter $\beta$ for the patient head data to achieve the best image quality;
specifically, we set $\beta$ as
$3000$ and $12000$ in testing BCD-Net-sCNN-lc and BCD-Net-sCNN-hc, respectively.

We trained a 100-iteration BCD-Net-dCNN, where we replaced image refining CNN architecture of BCD-Net-sCNN-hc with the aforementioned denoising dCNN architecture.  
We used the same training dataset used in training the non-iterative dCNN method.
We also used Adam optimization and identical settings (learning rate and regularization parameter $\beta$) as those of BCD-Net-sCNN-hc.
We set batch size and number of epochs as 1 and 10, respectively.
We observed with three test phantom samples that BCD-Net-dCNN becomes overfitted around 40th iteration; see Figure~\ref{fig:rmse-plot-10epoch}.  
We thus used the results at the 40th iteration for test phantom samples.
For the patient head data, we used 40-iteration BCD-Net-dCNN learned with XCAT phantom.
We set $\beta$ as 2400 after fine tuning to achieve the best image quality.

	\subsubsection{Evaluation metrics}
	In the quantitative evaluations with the XCAT phantom, we computed root-mean-square error (RMSE) for decomposed material images within a region of interest (ROI). 
	We set the ROI as a circle region that includes all the phantom tissue.
	For a decomposed material density image $\hat{\x}_m$, the RMSE in density (g/cm$^3$) is defined as $\sqrt{ \sum_{j=1}^{N_{\textup{ROI}}} (\hat{x}_{m,j}-x_{m,j}^\star)^2/{N_{\textup{ROI}}}}$,  where $x_{m,j}^\star$ denotes the true density value of the $m$th material at the $j$th pixel location, and $N_{\textup{ROI}}$ is the number of pixels in a ROI.  
	The ROI is indicated in red circle in Figure~\ref{fig:roi_anotate}(a).
	
	For the patient head data, we evaluated each method with \textit{1)} contrast-to-noise ratio (CNR) that measures the contrast between tissue of interest (TOI) and local background region, and \textit{2)} noise power spectrum (NPS)~\cite{zhu2015nsdect} that measures noise properties, in decomposed water images.
	CNR is defined as $\textup{CNR} = (\mu_{\textup{TOI}}-\mu_{\textup{BKG}}) / \sigma_{\textup{BKG}}$, where $\mu_{\textup{TOI}}$ and $\mu_{\textup{BKG}}$ are mean values in a TOI and local background region, respectively, and $\sigma_{\textup{BKG}}$ is standard deviation between pixel values in a local background region.
	We selected three TOI-local background sets in muscle and fat areas; see red and blue regions in Figure~\ref{fig:roi_anotate}(b).
	The NPS is defined as $\textup{NPS}=|\mathrm{DFT}\{f\}|^2$, where $f$ denotes the noise of a ROI of decomposed water image (the patient head data does not have the ground-truth, so we subtract the mean value  from the pixel values to approximate noise~\cite{zhu2015nsdect}), and $\mathrm{DFT}\{ \cdot \}$ applies the 2D discrete Fourier transform (DFT) to 2D image.
	We selected three ROIs with uniform intensity and of size $30\times 30$ in decomposed water image, and measured NPS within these ROIs; see the positions of three ROIs in Figure~\ref{fig:roi_anotate}(c).
	
	We used the most conventional measures for image quality assessment in tomography research. 
	In XCAT phantom experiments with available ground-truth material images, we calculated RMSE values for each method. In clinical data experiments, we used the CNR measure that is the most widely-used alternative to RMSE in tomography research particularly when ground-truths are unavailable.

	\subsection{Comparisons Between Different Methods with XCAT Phantom Data}
	\label{subsec:xcat_phantom}
	
	Table~\ref{Tab:rmse} summarizes the RMSE values of material images decomposed by different methods for three different test slices. 
	BCD-Net-sCNN-lc significantly decreases RMSE for material images compared to direct matrix inversion, DECT-EP, and DECT-ST.
	For all test samples, BCD-Net-sCNN-hc achieves significantly lower RMSE values compared to BCD-Net-sCNN-lc, implying the superiority of the distinct cross-material CNN architecture in~(\ref{eq:DisCro_Mapping}) over the identical encoding-decoding architecture in~(\ref{eq:IdInd_Mapping}).
	BCD-Net-sCNN-hc and dCNN methods achieve comparable errors:  BCD-Net-sCNN-hc achieves an average $0.4\times10^{-3}$\,g/cm$^3$ improvement for water images over dCNN, while dCNN achieves an average $0.2\times10^{-3}$\,g/cm$^3$ improvement for bone images over BCD-Net-sCNN-hc.
	Compared to BCD-Net-dCNN, BCD-Net-sCNN-hc gives higher average RMSE for bone images, and the same average RMSE for water images.		
	Compared to dCNN, BCD-Net-dCNN achieves RMSE improvements for both water and bone images, implying that dCNN denoisers combined with MBID modules in an iterative way can further decrease RMSE values.
	Figure~\ref{fig:average_rmse_plot} shows the RMSE convergence behavior of BCD-Net-sCNN-hc: it decreases monotonically. (See its fixed point convergence guarantee in the work~\cite{momen:19:il}.)	

	Figure~\ref{Fig:comp_xcat_decom} shows the \#1 material density images  of direct matrix inversion, DECT-EP, DECT-ST, dCNN, BCD-Net-sCNN-lc, BCD-Net-sCNN-hc, BCD-Net-dCNN, and ground truth. 
	DECT-EP reduces severe noise and artifacts in direct matrix inversion decompositions.
	DECT-ST, dCNN, and BCD-Net-sCNN-lc significantly improve the image quality  compared to DECT-EP, but still have some obvious artifacts.
	Compared to dCNN, BCD-Net-dCNN further reduces noise and artifacts and shows better recovery of the areas at the boundaries of water and bone; however, BCD-Net-dCNN still blurs soft-tissue regions.
	Compared to DECT-ST, dCNN, BCD-Net-sCNN-lc, and BCD-Net-dCNN, BCD-Net-sCNN-hc shows significantly better noise and artifacts reduction while improving the sharpness of edges in soft-tissue regions. 
	These improvements are clearly noticeable in the zoom-ins of water images.
	Decomposed material images for another two test slices are included in \crefrange{fig:test2}{fig:test3}. 

	\subsection{Comparisons Between Different Methods with Patient Data}
	\label{sec:clinical}
	
	Figure~\ref{fig:decom_head} shows decomposed material density images by different methods and high- and low-energy attenuation images for clinical head data. 
	DECT-EP reduces severe noise and artifacts  in direct matrix inversion results, but  it is difficult to distinguish edges in many soft tissue regions.
	DECT-ST and dCNN suppress noise and improve the edges in soft tissues compared to DECT-EP, but both still have poor contrast in many soft tissue regions. 
	BCD-Net-sCNN-lc and BCD-Net-dCNN further improve the contrast in soft tissue regions compared to DECT-ST and dCNN. 
	However, BCD-Net-sCNN-lc has bright artifacts---see the bottom-right zoom-in in water image---and BCD-Net-dCNN leads to indistinguishable bone marrow structures---see the bottom-left zoom-ins in water and bone images.
	BCD-Net-sCNN-hc better removes noise and artifacts, provides clearer image edges and structures, and recovers subtle details, compared to the other methods aforementioned.
	One clearly noticeable improvement is captured in the bottom-right zoom-ins in water images, where BCD-Net-sCNN-hc not only improves edge sharpness and contrast in soft tissue, but also suppresses bright artifacts.
	Inside the red circle 1 in water images, BCD-Net-sCNN-hc and BCD-Net-dCNN preserve a ``dark spot"  that exists in attenuation images,  whereas DECT-EP, DECT-ST, dCNN, and BCD-Net-sCNN-lc all missed it.  
	The structure of the dark spot is an artery that contains diluted iodine solution caused by angiogram.
	The linear attenuation coefficient of iodine is much closer to bone than soft-tissue. During decomposition, most of the iodine is grouped into the bone image, while in the water image there are only some pixels with tiny values, thus it is a dark spot.
	Moreover, the marrow structures obtained by BCD-Net-sCNN-hc have sharper edges (inside red circle 2) than the other methods.

	Table~\ref{tab:cnr} summaries the CNR values for the three different TOI-local background sets in the decomposed water images via different methods.
	BCD-Net-sCNN-hc achieves significantly higher CNR compared to the other methods for all the three TOI-local background sets, and the performance degrades in the following order: BCD-Net-dCNN, BCD-Net-sCNN-lc, dCNN, DECT-ST, DECT-EP, direct matrix inversion.
	In particular, BCD-Net-sCNN-hc achieves 1.70 improvement in CNR in average over BCD-Net-dCNN, and BCD-Net-dCNN achieves 3.14 improvement in CNR in average over dCNN.

	Figure~\ref{fig:nps} compares the magnitude of NPS from different methods.
	Across all  frequencies, the NPS magnitude of BCD-Net-sCNN-hc is significantly smaller than that of direct matrix inversion, DECT-EP, DECT-ST, and dCNN.
	The overall low-frequency noise of BCD-Net-sCNN-hc is also significantly less than that of the aforementioned methods.
	What is more, BCD-Net-sCNN-hc achieves fewer vertical and horizontal frequency strips with lower intensity compared to BCD-Net-sCNN-lc and BCD-Net-dCNN, especially in the ROI \#1 and \#3.
	The aforementioned NPS comparisons demonstrate the superiority of the proposed BCD-Net-sCNN-hc in  removing noise and artifacts inside soft tissue regions.
	We observed similar trends in averaged NPS measures using multiple noise realizations; see Figure~\ref{fig:nps-xcat}.

	Similar to XCAT phantom results, the dCNN denoiser and BCD-Net-dCNN give less appealing material images of the clinical head data, compared to the proposed BCD-Net-sCNN-hc.
	We conjecture that the following reasons may limit the dCNN denoising performance:
	lack of considering decomposition physics and/or limited training samples and diversity.
	Although BCD-Net-dCNN incorporates decomposition physics, due to too high NN complexity (compared to the diversity of the training data), the image quality for both phantom and patient head data are still unsatisfactory.
	The proposed method, BCD-Net-sCNN-hc, resolves the issues of dCNN and BCD-Net-dCNN by using both MBID cost minimization and shallow CNN refiner at each iteration.
	The clinical head data shows that the proposed BCD-Net-sCNN-hc successfully reduces noise/artifacts and preserves subtle details that exist in attenuation images in Figure~\ref{fig:decom_head}.

	\subsection{Computational Complexity Comparisons}
	\label{sec:comp}
	The computational cost of DECT-EP, DECT-ST, and the proposed BCD-Net-sCNNs scale as $O(R_{\textup{EP}}N I_{\textup{EP}})$, $O((R_\textup{ST})^2N I_{\textup{ST}})$, and $O(RKNI_\mathrm{iter})$, respectively, 
	where $I_{\textup{EP}}$ and $I_{\textup{ST}}$ are the number of iterations for DECT-EP and DECT-ST, respectively. 
	The computational cost of the chosen dCNN architecture in Section~\ref{sec:dcnn-denoiser} and BCD-Net-dCNN  scale as $O(R_{\textup{dCNN}} K_\textup{dCNN} N ( (C-2)K_\textup{dCNN}+4 ))$ and $O(R_{\textup{dCNN}} K_\textup{dCNN} N ( (C-2)K_\textup{dCNN}+4 )I_\textup{dCNN})$, respectively,
	where $R_\textup{dCNN}$, $K_{\textup{dCNN}}$, and $C$ are 
	kernel size, the number of features, and the number of convolutional layers of dCNN denoiser, respectively, and $I_\textup{dCNN}$ is the number of BCD-Net-dCNN iterations.
	In all experiments, we used $R_{\textup{EP}}=8$ and $I_{\textup{EP}}=500$ for DECT-EP, $R_\textup{ST}=64$ and $I_{\textup{ST}}=1000$ for DECT-ST, $R_\textup{dCNN}=3^2$, $K_{\textup{dCNN}}=64$, and $C=4$ for dCNN denoiser, $I_\textup{dCNN}=40$ for BCD-Net-dCNN, and $R=K=8^2$ and $I_\mathrm{iter}=100$ for the proposed BCD-Net-sCNN-hc.
	The big-O analysis reveals that the computational cost of 100-iteration of the proposed BCD-Net-sCNN-hc is larger than 500-iteration DECT-EP and the chosen dCNN denoiser, $87\%$ cheaper than that of 40-iteration BCD-Net-dCNN, and $90\%$ cheaper than that of 1000-iteration DECT-ST.

	\subsection{Discussions for Generalization Performance of dCNN, BCD-Net-dCNN, and BCD-Net-sCNN-hc}
	To study the generalization performance of dCNN, BCD-Net-dCNN, and BCD-Net-sCNN-hc, we calculated the average RMSE values from training and test samples, and their difference.
	Table~\ref{tab:rmse_gap} reports the RMSE gap between decomposed images in training and test via dCNN, BCD-Net-dCNN, and BCD-Net-sCNN-hc.
	BCD-Net-dCNN has smaller RMSE gap for both water and bone images, compared to dCNN that lacks decomposition physics.
	We conjecture that including MBID modules in an iterative way can improve the generalization performance of dCNN denoisers.
	This result is well aligned with the recent work~\cite{wu:21:ar} demonstrating that combining deep NNs, imaging physics, and sparisty-promoting regularizer gives the stable performance against perturbations.
	BCD-Net-sCNN-hc has smaller RMSE gap for both water and bone images, compared to BCD-Net-dCNN.
	At each BCD-Net iteration, the number of trainable parameters are $2K(4R+1)$ and $R_{\textup{dCNN}} K_\textup{dCNN}( (C-2)K_\textup{dCNN}+4 )$ for BCD-Net-sCNN-hc and BCD-Net-dCNN, respectively;
	specifically, they are 32,896 and 76,032 using the parameter sets in Section~\ref{sec:comp}.
	We conjecture that sCNN-hc refiner with lower NN complexity can improve the generalization performance over dCNN refiner.

	\section{Conclusions}
	\label{sec:conclusions}
	Image-domain decomposition methods are readily applicable to commercial DECT scanners, but susceptible to noise and artifacts on attenuation images.
	To improve MBID performance,  it is important to incorporate accurate prior knowledge into sophisticatedly designed MBID.
	The proposed INN architecture, BCD-Net-sCNN-hc, successfully achieves accurate MBID by providing accurate prior knowledge via its iteration-wise refiners that exploit correlations between different material images with distinct encoding-decoding filters.
	Our study with patch-based reformulation reveals that learned filters of distinct cross-material CNN refiners can approximately satisfy the tight-frame condition and useful for noise suppression and signal restoration.
	On both XCAT phantom and patient head data, the proposed BCD-Net-sCNN-hc  reduces the artifacts at boundaries of materials and improves edge sharpness and contrast in soft tissue, compared to 
	a conventional MBID method, DECT-EP, a recent unsupervised MBID method, DECT-ST, and a noniterative dCNN method.
	We also show that BCD-Net-sCNN-hc improves the image quality over BCD-Net-dCNN, especially for patient head data, potentially due to its lower refiner complexity over that of BCD-Net-dCNN.
	For choosing refiner architecture in BCD-Net, we suggest considering the number of trainable parameters with the size/diversity of training data.
		
	There are a number of avenues for future work.
	Our first future work is to investigate a three-material decomposition BCD-Net architecture in DECT; see its potential benefit in Section~\ref{sec:result-supp} and \crefrange{fig:mmd}{fig:comp-two-three}.
	Second, to further improve the MBID model, we plan to train the weight matrix $\W_0$ in (\ref{eq:MBID}) in a supervised way with proper loss function designs, rather than statistically estimating it.
	By extending the patch-perspective interpretations, we will develop an ``explainable" deeper refiner that might further improve the MBID performance of BCD-Net.
	Third, to accommodate the non-trivial tuning process of $\beta$ in (\ref{eq:MBID}), we plan to learn it from training datasets.
	Finally, to further improve the generalization capability of the proposed INN architecture, we will additionally incorporate a sparsity-promoting regularizer into the proposed framework, similar to the recent work~\cite{wu:21:ar}.
	
	\section{Acknowledgement}
	The authors thank Dr. Tianye Niu, Shenzhen Bay Laboratory, for providing clinical DECT images for our experiments.
	
	\section{Conflict of Interest Statement}
	The authors have no relevant conflicts of interest to disclose.

	\section{Data Availability}
	The data that support the findings of this study are available from the corresponding author upon reasonable request.

	\section*{References}
	\addcontentsline{toc}{section}{\numberline{}References}
	\vspace*{-10mm}
	\bibliography{./refs_bcdnet}      
	
	\makeatletter\@input{yy.tex}\makeatother
	
	\clearpage
	
	\begin{figure}[!t]
		\centering
		\includegraphics[scale=1.3]{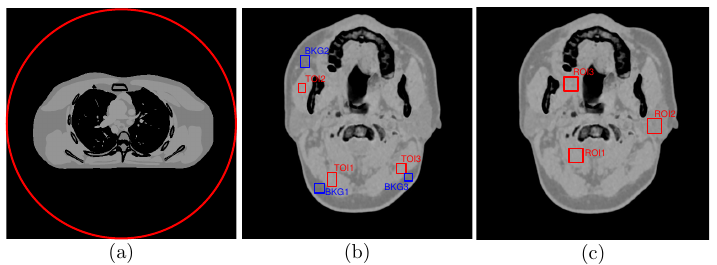}
		\vspace{-6pt}
		\caption{(a) ROI used for RMSE calculation for XCAT phantom data. (b) Three selected TOIs in muscle (indicated by red rectangles) and corresponding local background regions in fat (indicated by blue rectangles) on the decomposed water image of head data. (c) Three selected ROIs for NPS calculation for the decomposed water image of head data. }
		\label{fig:roi_anotate}
	\end{figure}

	\begin{figure}[!t]
		\centering
		\includegraphics[scale=0.6]{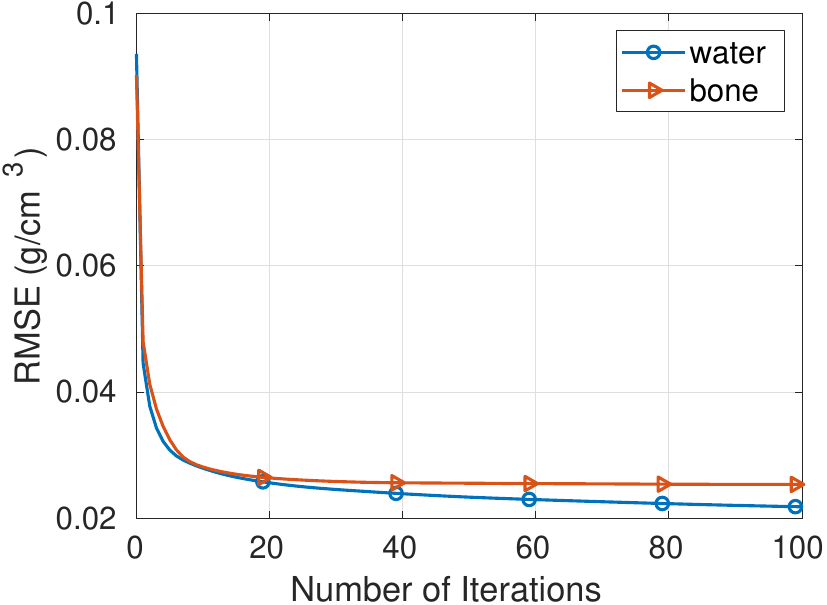}
		\vspace{-6pt}
		\caption{RMSE convergence behaviors of BCD-Net-sCNN-hc (averaged RMSE values across three test slices of XCAT phantom).}
		\label{fig:average_rmse_plot}
		\vspace{-6pt}
	\end{figure}

	\begin{figure*}[!t] 
		\centering
		\includegraphics[scale=1.35]{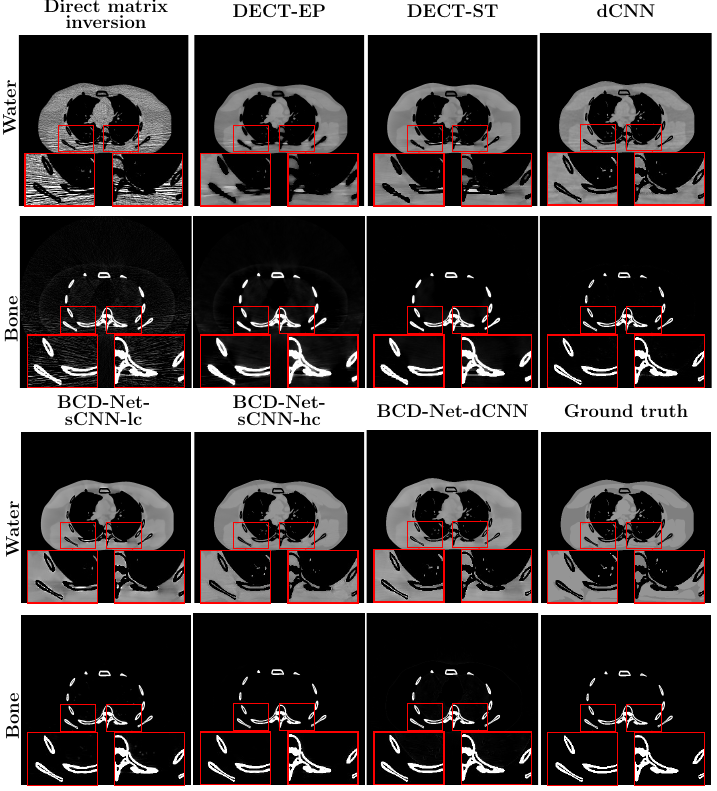}
		\vspace{-8pt}
		\caption{Comparison of decomposed images from different methods (XCAT phantom test slice \#1). Water and bone images are shown with display windows [0.7  1.3]\,g/cm$^3$ and [0 0.8]\,g/cm$^3$, respectively. }		
		\label{Fig:comp_xcat_decom}
	\end{figure*}

	\begin{figure*}[!t]
		\centering
		\includegraphics[scale=1.35]{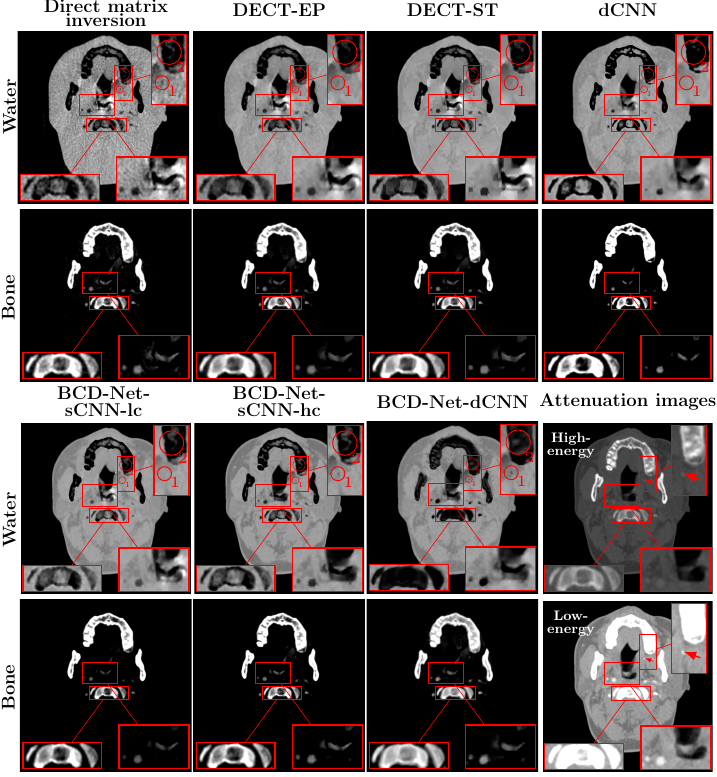}
		\vspace{-8pt}
		\caption{Comparison of decomposed images from different methods (clinical head data). 
			Water and bone images are displayed with windows [0.5 1.3]~g/cm$^3$ and [0.05 0.905]~g/cm$^3$, respectively. 
			High- and low-energy attenuation images are displayed with window [0.1 0.35]\,cm$^{-1}$.}
		\label{fig:decom_head}
		\vspace{0pt}
	\end{figure*}	

	\begin{figure*}[!t]
		\centering
		\includegraphics[scale=1.38]{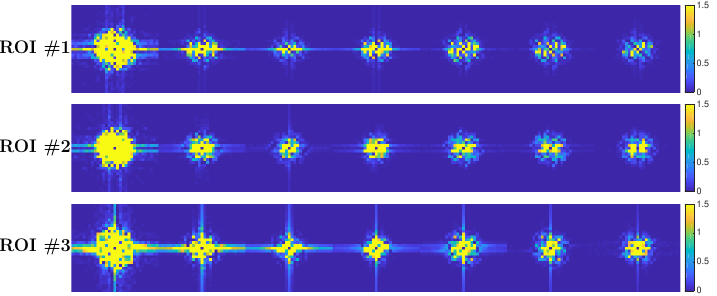}
		\vspace{-8pt}
		\caption{Left to right: NPS measured within ROIs of decomposed water images obtained by direct matrix inversion, DECT-EP, DECT-ST, dCNN, BCD-Net-dCNN, BCD-Net-sCNN-lc, and BCD-Net-sCNN-hc. 
			The first to the third rows show the NPS of the first to third ROI in Figure~\ref{fig:roi_anotate}(c), respectively, with display windows [0 1.5]~g$^2$/cm$^6$.}
		\label{fig:nps}
	\end{figure*}

\clearpage

	\begin{table}[!t]
		\centering
		\caption{Data acquisition parameters applied in head data acquisition.}
		\vspace{-6pt}
		\begin{tabular}{p{7.5cm}<{\centering}  p{3cm}<{\centering} p{3cm}<{\centering}}
			\hline\hline
			\multirow{2}{*}{Scanner}  & \multicolumn{2}{c}{Head Data}
			\\	\cline{2-3}
			& \tabincell{c}{High-energy \\}
			&  \tabincell{c}{Low-energy \\} \\ \hline  
			Peak Voltage (kVp) & 140 & 80 \\ \hline
			X-ray Tube Current (mA) & 364 & 648 \\ \hline
			Exposure Time (s) & \multicolumn{2}{c}{0.285} \\ \hline
			Current-exposure Time Product (mAs) & 103.7 & 184.7 \\ \hline
			Noise STD (mm$^{-1}$) & $1.57 \times 10^{-4}$ & $3.61\times10^{-4}$ \\ \hline
			Helical Pitch & \multicolumn{2}{c}{0.7}\\ \hline
			Gantry Rotation Speed (circle/second) & \multicolumn{2}{c}{0.28} \\ 
			\hline	\hline				   
		\end{tabular} 
		\label{Tab:clinical}	
		\vspace{0.1in}
	\end{table}

	\begin{table}[!t]
		\centering
		\caption{RMSE of decomposed material density images obtained by different methods for three different test slices of XCAT phantom. 
			The unit for RMSE is $10^{-3}$\,g/cm$^3$. }
		\vspace{-6pt}
		\begin{tabular}{ p{4cm}<{\centering} p{1cm}<{\centering} p{1cm}<{\centering} p{1cm}<{\centering} p{1cm}<{\centering} p{1cm}<{\centering} p{1cm}<{\centering} p{1cm}<{\centering} p{1cm}<{\centering}}
			\hline\hline
			\multirow{2}{*}{{Methods}}  & \multicolumn{2}{c}{Test \#1} & \multicolumn{2}{c}{Test \#2} & \multicolumn{2}{c}{Test \#3} & \multicolumn{2}{c}{Average}\\ \cline{2-9}
			& water & bone & water & bone & water & bone  & water & bone \\ \hline
			Direct matrix inversion & 91.2 & 89.0 & 70.4 & 69.9 & 119.2 & 111.9 & 93.6 & 90.3 \\ \hline
			DECT-EP & 60.0 & 68.5 & 59.5 & 63.3 & 69.9 & 75.9 & 63.1 & 69.2\\ \hline
			DECT-ST & 54.2 & 60.3 & 52.1 & 54.1 & 62.5 & 66.3 & 56.3 & 60.2\\ \hline
			\tabincell{c}{dCNN} & 21.9 & 24.3 & 19.8 & 20.8 & 24.9 & 30.2 & 22.2 & 25.1\\  \hline
			\tabincell{c}{BCD-Net-sCNN-lc} & 44.4 & 39.1 & 37.0 & 33.4 & 47.2 & 48.8 & 42.9 & 40.4\\ \hline
			\tabincell{c}{BCD-Net-sCNN-hc}  & 23.0 & 25.3 & 20.2 & 23.2 & 22.2 & 27.6 & \textbf{21.8} & 25.3 \\ \hline 
			\tabincell{c}{BCD-Net-dCNN}  & 22.7 & 23.4 & 22.0 & 22.6 & 20.7 & 22.0 & \textbf{21.8} & \textbf{22.7}
			\\
			\hline \hline  
		\end{tabular} 
		\label{Tab:rmse}
		\vspace{-6pt}
	\end{table}

	\begin{table}[!t]
		\centering
		\caption{CNR of decomposed water density images obtained by different methods for clinical head data.}
		\vspace{-6pt}
		\begin{tabular}{ p{5cm}<{\centering} p{2.5cm}<{\centering} p{2.5cm}<{\centering} p{2.5cm}<{\centering} p{2cm}<{\centering}}
			\hline\hline
			& \tabincell{c}{TOI-local \\[-0.9em] BKG \#1} & \tabincell{c}{TOI-local \\[-0.9em] BKG \#2} & \tabincell{c}{TOI-local \\[-0.9em] BKG \#3} & \tabincell{c}{Average}\\ \hline
			Direct matrix inversion & -0.05 & -0.21 & 0.05 & -0.06\\ \hline
			DECT-EP & 0.14 & -0.28 & 0.63 & 0.16\\ \hline
			DECT-ST & 1.97 & 0.18 & 3.44 & 1.86 \\ \hline
			\tabincell{c}{dCNN} & 5.08 & 4.92 & 4.46 & 4.82\\  \hline
			\tabincell{c}{BCD-Net-sCNN-lc} & 6.83 & 8.45 & 5.39 & 6.89 \\ \hline
			\tabincell{c}{BCD-Net-sCNN-hc}  & 10.01 & 11.48 & 7.49 & \textbf{9.66}\\ \hline
			\tabincell{c}{BCD-Net-dCNN}  & 8.16 & 9.44 & 6.29 & 7.96 \\ 
			\hline \hline  
		\end{tabular} 
		\label{tab:cnr}
		\vspace{8pt}
	\end{table}

	\begin{table}[!t]
		\centering
		\caption{RMSE of decomposed density images from training and test samples via dCNN, BCD-Net-dCNN, and BCD-Net-sCNN-hc. RMSE gap is the difference between test RMSE and training RMSE.
			The unit for RMSE is $10^{-3}$\,g/cm$^3$. }
		\begin{tabular}{ p{1cm}<{\centering}   p{1.5cm}<{\centering} p{1.5cm}<{\centering} p{1.5cm}<{\centering}  p{1.5cm}<{\centering} p{1.5cm}<{\centering} p{2.cm}<{\centering} p{2.cm}<{\centering}}
			\hline\hline
			&	\tabincell{c}{Methods}  & \multicolumn{2}{c}{\tabincell{c}{dCNN}} & \multicolumn{2}{c}{\tabincell{c}{BCD-Net-dCNN}} & \multicolumn{2}{c}{\tabincell{c}{BCD-Net-sCNN-hc}} \\ \hline
			& &  water & bone  & water & bone & water & bone 
			\\\hline
			\multirow{3}{*}{\tabincell{c}{RMSE}} & Training & 18.4 & 21.6 & 18.7 & 19.4 & 21.5 & 22.8    \\
			& Test & 22.2 & 25.1 & 21.8 & 22.7 & 21.8 & 25.4  \\
			& Gap &  3.8 & 3.5 & 3.1 & 3.3 & 0.3  &  2.6 
			\\\hline
			\hline
		\end{tabular} 
		\label{tab:rmse_gap}
	\end{table}	

\clearpage
\large{\textbf{List of Figures:}}
\begin{itemize}
	\item Figure~1: The proposed BCD-Net architecture at the $i$th iteration, for $i=1,\ldots,I_{\text{iter}}$.
	\item Figure~2: $\D^{(100)} \E^{(100)}$ of BCD-Net-sCNN-hc.
	\item Figure~3: Left and right are learned filters of BCD-Net-sCNN-lc and BCD-Net-sCNN-hc at the last iteration that uses identical encoding-decoding architecture (i.e., $\D=\E^\top$), respectively.
	Top-left, top-right, bottom-left, and bottom-right correspond to $\E_{1,1}$, $\E_{1,2}$, $\E_{2,1}$, and $\E_{2,2}$, respectively. 
	Four pairs of filters (indicated by four different colors) are selected as examples to show similar or different structures between off-diagonal and diagonal blcok matrices;
	filters indicated by red or green boxes show similar structures, while blue or yellow boxes show different structures.
	\item Figure~4: The attenuation images (zoomed-in) for a test slice at high and low energies, respectively.
	\item Figure~5: (a) ROI used for RMSE calculation for XCAT phantom data. (b) Three selected TOIs in muscle (indicated by red rectangles) and corresponding local background regions in fat (indicated by blue rectangles) on the decomposed water image of head data. (c) Three selected ROIs for NPS calculation for the decomposed water image of head data. 
	\item Figure~6: RMSE convergence behaviors of BCD-Net-sCNN-hc (averaged RMSE values across three test slices of XCAT phantom).
	\item Figure~7: Comparison of decomposed images from different methods (XCAT phantom test slice \#1). Water and bone images are shown with display windows [0.7  1.3]\,g/cm$^3$ and [0 0.8]\,g/cm$^3$, respectively.
	\item Figure~8: Comparison of decomposed images from different methods (clinical head data). Water and bone images are displayed with windows [0.5 1.3]~g/cm$^3$ and [0.05 0.905]~g/cm$^3$, respectively. 
	High- and low-energy attenuation images are displayed with window [0.1 0.35]\,cm$^{-1}$.
	\item Figure~9:	Left to right: NPS measured within ROIs of decomposed water images obtained by direct matrix inversion, DECT-EP, DECT-ST, dCNN, BCD-Net-dCNN, BCD-Net-sCNN-lc, and BCD-Net-sCNN-hc. 
	The first to the third rows show the NPS of the first to third ROI in Figure~\ref{fig:roi_anotate}(c), respectively, with display windows [0 1.5]~g$^2$/cm$^6$.
	\item Figure~S.1: RMSE plot of BCD-Net-dCNN for Test \#1, Test \#2, and Test \#3, respectively.
	\item Figure~S.2: (a) Five selected ROIs indicated for $\overline{\textup{NPS}}$ calculation for the decomposed water image of XCAT phantom. (b) Left to right: NPS measured within ROIs of decomposed water images obtained by direct matrix inversion, DECT-EP, DECT-ST, dCNN, BCD-Net-dCNN, BCD-Net-sCNN-lc, and BCD-Net-sCNN-hc.
	The first to the fifth rows in (b) show the $\overline{\textup{NPS}}$ of the first to fifth ROIs, respectively, with display windows [0 0.6]~g$^2$/cm$^6$.
	\item Figure~S.3: Comparison of decomposed images from different methods (XCAT phantom test slice \#2). Water and bone images are shown with display windows [0.7  1.3]\,g/cm$^3$ and [0 0.8]\,g/cm$^3$, respectively.
	\item Figure~S.4: Comparison of decomposed images from different methods (XCAT phantom test slice \#3).  
	Water and bone images are displayed with windows [0.7  1.3]\,g/cm$^3$ and [0 0.8]\,g/cm$^3$, respectively.
	\item Figure~S.5: Comparison of three decomposed images from regularized direct matrix inversion ($\lambda=1\times 10^{-5}$), BCD-Net-sCNN-hc, and ground truth. 
	Fat, muscle, and bone images are shown with display windows [0 2]~g/cm$^3$, [0 2]~g/cm$^3$, and [0 0.5]~g/cm$^3$, respectively.
	\item Figure~S.6: RMSE convergence behaviors of three-material decomposition BCD-Net-sCNN-hc.
	\item Figure~S.7: Comparisons of decomposed bone images (display window [0 0.5]~g/cm$^3$) and their error maps (display window [0 0.3]~g/cm$^3$) from dual- and three-material decomposition BCD-Net-sCNN-hc architectures.	
\end{itemize}

\end{document}


\cen{\sf {\Large {\bfseries  An Improved Iterative Neural Network for High-Quality Image-Domain Material Decomposition in Dual-Energy CT -- Supplementary Material} \\  
\vspace{10mm}
}
\vspace{-10mm}
}

\pagenumbering{roman}
\setcounter{page}{1}
\pagestyle{plain}

\setlength{\baselineskip}{0.7cm}      

\pagenumbering{arabic}
\setcounter{page}{1}
\pagestyle{fancy}

This supplement provides 
details for optimizing the training loss function in (\ref{eq:DisCro_Loss}), relation between convolution-perspective and patch-based trainings for distinct cross-material CNN refiner in (\ref{eq:DisCro_Mapping}), and additional experimental results to accompany our main manuscript~\cite{bcdnet:20:li}.
We use the prefix ``S”  for the numbers in section, proposition, equation, and figure in the supplementary material.
\setcounter{section}{0}
\setcounter{figure}{0}
\setcounter{table}{0}
\setcounter{equation}{0}

\section{Optimizing  (\ref{eq:DisCro_Loss}) with a Mini-Batch Stochastic Gradient Method}
\label{sec:subdiff}
The training loss at each mini-batch is
\begin{equation*}
\label{eq:rewrite_cost}
\begin{split}
\mathcal{L}=&\frac{1}{B} \sum_{r=1}^{2R} \sum_{b=1}^{B} \left(X_{rb}-\D_r \TT_{\exp(\alp)} \left(\E \X_{b}^{(i-1)}\right) \right)^2 \\
=&\sum_{r=1}^{2R} \sum_{b=1}^{B} \frac{1}{B} \left[X_{rb}-\D_r \left(\sum_{r=1}^{2R}\E_r X_{rb}^{(i-1)}-\exp(\alp) \odot \right.\right.  \left.\left. \textup{sign}\left(\sum_{r=1}^{2R} \E_r X_{rb}^{(i-1)}\right) \right)\odot \mathbbm{1}_{|\E \X_b^{(i-1)}|>\exp(\alp)} \right]_,^2
\end{split}
\end{equation*}
where $\D_r$ is the $r$th row of $\D$, $\E_r$ is the $r$th column of $\E$,  $(\cdot)_{rb}$ denotes the element at $r$th row and $b$th column of the matrix. 
Therefore, subgradient of (\ref{eq:DisCro_Loss}) with respect to $\alp$ at each mini-batch is

\begin{equation*}
\label{eq:derivation_gradgamma}
\begin{split}
&\frac{\partial \mathcal{L}(\D,\,\E,\,\alp)}{\partial \alp} = \frac{2}{B} \sum_{r=1}^{2R} \sum_{b=1}^{B} \left[X_{rb}-\D_r\TT_{\exp(\alp)}\left(\E\X_b^{(i-1)}\right) \right] \cdot \\
&\frac{\partial \D_r \left[\exp(\alp)\odot \textup{sign}\left(\sum_{r=1}^{2R}\E_r X_{rb}^{(i-1)}\right) \right]\odot \mathbbm{1}_{|\E\X_b^{(i-1)}|>\exp(\alp)} }{\partial \alp}  \\
=&\frac{2}{B} \sum_{r=1}^{2R} \sum_{b=1}^{B} \left[X_{rb}-\D_r\TT_{\exp(\alp)}\left(\E\X_b^{(i-1)}\right) \right] \cdot \D_r^\top\odot  \exp(\alp) \odot \textup{sign} \left(\E\X_b^{(i-1)}\right) \odot \mathbbm{1}_{|\E\X_b^{(i-1)}|>\exp(\alp)} \\
=& \frac{2}{B} \left\{\D^\top\left(\X-\D\Z^{(i-1)} \right) \odot \exp(\alp\mathbf{1}') \odot \textup{sign}\left(\E\X^{(i-1)}\right) \odot \right.  \left.  \mathbbm{1}_{|\E\X^{(i-1)}|>\exp(\alp\mathbf{1}')} \right\}\mathbf{1} \\
=& \frac{2}{B} \left\{\D^\top\left(\X-\D\Z^{(i-1)} \right) \odot \exp(\alp\mathbf{1}') \odot \textup{sign}\left(\Z^{(i-1)}\right) \right\}\mathbf{1}_.
\end{split}
\end{equation*}

We can easily obtain subgradient of $\mathcal{L}$ with respect to $\D$ at each mini-batch as
\begin{equation*}
\label{eq:derivation_gradD}
\frac{\partial \mathcal{L}}{\partial \D} = -\frac{2}{B} \left(\X-\D \Z^{(i-1)}\right)\cdot \Z^{(i-1)^\top}.
\end{equation*} 

At each mini-batch,  the subgradient of $\mathcal{L}$ with respect to $r_1$th column of $\E$ is as follows:
\begin{equation*}
\begin{aligned}
&\frac{\partial \mathcal{L}(\D,\E,\alp)}{\partial \E_{r_1}}  
=-\frac{2}{B}\sum_{r=1}^{2R}\sum_{b=1}^{B} \left(X_{rb}-\D_r\TT_{\exp(\alp)}\left(\E\X_b^{(i-1)}\right) \right) \cdot   \quad \D_r^\top \odot   \mathbbm{1}_{|\E\X_b^{(i-1)}|>\exp(\alp)} \cdot X_{r_1b}^{(i-1)} \\ 
=&-\frac{2}{B} \sum_{b=1}^{B} \D^\top\left(\X_b - \D \TT_{\exp(\alp)}\left(\E\X_b^{(i-1)}\right) \}\right)  \odot  \mathbbm{1}_{|\E\X_b^{(i-1)}|>\exp(\alp)} \cdot X_{r_1b}^{(i-1)} \\
=&-\frac{2}{B} \D^\top\left(\X-\D \Z^{(i-1)}\right) \odot \mathbbm{1}_{|\E\X^{(i-1)}|>\exp(\alp\mathbf{1}')} \cdot {\X_{r_1}^{(i-1)}}^\top.
\end{aligned}
\end{equation*}
Thus, the subgradient of $\mathcal{L}$ with respect to $\E$ for each mini-batch selection is
\begin{equation*}
\label{eq:derivation_gradW}
\begin{split}
\frac{\partial \mathcal{L}(\D,\,\E,\,\alp)}{\partial \E} = &-\frac{2}{B} \D^\top\left(\X - \D\TT_{\exp{(\alp\mathbf{1'})}}\left(\E\X^{(i-1)}\right)\right) \odot  \mathbbm{1}_{|\E\X^{(i-1)}|>\exp{(\alp\mathbf{1'})}} \cdot {\X^{(i-1)}}^\top 
\end{split}
\end{equation*}

\section{Relation between convolution-perspective and patch-based trainings of the proposed BCD-Net-sCNN-hc}
\label{sec:patch-conv}

\prop{
	\label{prop:refiner}
	The proposed CNN refiner in (\ref{eq:DisCro_Mapping}) can be rewritten with patch-based perspective as follows (we omit the iteration superscript indices $(i)$ for simplicity):
	\begin{equation}
	\left[    \hspace{-3pt}
	\begin{array}{c}
	\sum_{k=1}^{K}\sum_{n=1}^{2} \dd_{1,n,k} *\TT_{\exp({\alpha_{n,k}})} \left( \sum_{m=1}^{2} \ee_{n,m,k} * \x_{m} \right) \\
	\sum_{k=1}^{K}\sum_{n=1}^{2} \dd_{2,n,k} *\TT_{\exp({\alpha_{n,k}})} \left( \sum_{m=1}^{2} \ee_{n,m,k} * \x_{m} \right)
	\end{array}    \hspace{-3pt}
	\right] = \frac{1}{R}\sum_{j=1}^{N} \bar{\P}_j^\top \D \TT_{\exp{(\alp)}} (\E\bar{\P}_j \x),
	\end{equation}
	where $\x=[\x_1^\top,\x_2^\top]^\top$. See other related notations in (\ref{eq:DisCro_Mapping}) and (\ref{eq:conv2patch}).}

\noindent\textit{Proof.} First, we have the following reformulation~\cite{CAOL:19:il:supp}: 
\begin{equation*}
\left[
\begin{array}{c}
\ee_{n,m,1} * \u\\ \vdots \\ \ee_{n,m,K} * \u
\end{array}
\right]=
\P\left[
\begin{array}{c}
\E_{n,m}\P_1\\ \vdots \\\E_{n,m}\P_{N}
\end{array}
\right]\u := \widetilde{\E}_{n,m} \u,
\end{equation*}
where $\P\in \mathbb{R}^{KN\times KN}$ is a permutation matrix. Considering that
\begin{equation*}
\sum_{k=1}^{K} \bar{\ee}_{n,m,k}*\left(\ee_{n,m,k}*\u \right)=\frac{1}{R}\widetilde{\E}_{n,m}^H \widetilde{\E}_{n,m} \u, 
\end{equation*}
we have 
\begin{equation*}
\sum_{k=1}^{K} \dd_{1,1,k}*\left(\ee_{1,1,k}*\x_{1} \right)=\frac{1}{R}\widetilde{\D}_{1,1} \widetilde{\E}_{1,1} \x_{1}
\quad
\textup{and} 
\quad
\sum_{k=1}^{K} \dd_{1,1,k}*\left(\ee_{1,2,k}*\x_{2} \right)=\frac{1}{R}\widetilde{\D}_{1,1} \widetilde{\E}_{1,2} \x_{2}.
\end{equation*}
Then we obtain the following reformulation result for term $ \sum_{k=1}^{K} \dd_{1,1,k}*\TT_{\exp{(\alpha_{1,k})}}(\ee_{1,1,k}*\x_{1}+\ee_{1,2,k}*\x_{2} )$: 
\begin{equation}
\centering
\sum_{k=1}^{K} \dd_{1,1,k}*\TT_{\exp{(\alpha_{1,k})}}\left(\ee_{1,1,k}*\x_{1}+\ee_{1,2,k}*\x_{2} \right)=
\frac{1}{R}\sum_{j=1}^{N} \P_j^\top \D_{1,1} \TT_{\exp{(\alp_1)}}\left( \E_{1,1}\P_j \x_{1} +\E_{1,2}\P_j\x_{2}\right),
\label{eq:d11}
\end{equation}
where we use the permutation invariance of thresholding operator~\cite{momen:19:il:supp} and $\P^\top\P = \I$. 
Similarly, for term $ \sum_{k=1}^{K} \dd_{1,2,k}*\TT_{\exp{(\alpha_{2,k})}}(\ee_{2,1,k}*\x_{1}+\ee_{2,2,k}*\x_{2})$, we have 
\begin{equation}
\centering
\sum_{k=1}^{K} \dd_{1,2,k}*\TT_{\exp{(\alpha_{2,k})}}\left(\ee_{2,1,k}*\x_{1}+\ee_{2,2,k}*\x_{2} \right)=
\frac{1}{R}\sum_{j=1}^{N} \P_j^\top \D_{1,2} \TT_{\exp{(\alp_2)}}\left( \E_{2,1}\P_j \x_{1} +\E_{2,2}\P_j\x_{2}\right).
\label{eq:d12}
\end{equation}
Combining \labelcref{eq:d11,eq:d12} gives the following result: 
\begin{equation}
\centering
\label{eq:d1n}
\begin{split}
\sum_{k=1}^{K}\sum_{n=1}^{2} \dd_{1,n,k} *\TT_{\exp{(\alpha_{n,k})}} \left( \sum_{m=1}^{2} \ee_{n,m,k} * \x_{m} \right) = \frac{1}{R}\sum_{j=1}^{N} \P_j^\top \D_{1,1} \TT_{\exp{(\alp_1)}} \left( \E_{1,1}\P_j\x_{1}+\E_{1,2}\P_j\x_{2}\right) + \\ \frac{1}{R}\sum_{j=1}^{N} \P_j^\top\D_{1,2} \TT_{\exp{(\alp_2)}} \left( \E_{2,1}\P_j\x_{1}+\E_{2,2}\P_j\x_{2}\right). 
\end{split}
\end{equation}
Similar to (\ref{eq:d1n}), we have 
\begin{equation}
\centering
\label{eq:d2n}
\begin{split}
\sum_{k=1}^{K}\sum_{n=1}^{2} \dd_{2,n,k} *\TT_{\exp{(\alpha_{n,k})}} \left( \sum_{m=1}^{2} \ee_{n,m,k} * \x_{m} \right) = \frac{1}{R}\sum_{j=1}^{N} \P_j^\top \D_{2,1} \TT_{\exp{(\alp_1)}} \left( \E_{1,1}\P_j\x_{1}+\E_{1,2}\P_j\x_{2}\right) + \\ \frac{1}{R}\sum_{j=1}^{N} \P_j^\top\D_{2,2} \TT_{\exp{(\alp_2)}} \left( \E_{2,1}\P_j\x_{1}+\E_{2,2}\P_j\x_{2}\right). 
\end{split}
\end{equation}
Combining the results in (\ref{eq:d1n}) and (\ref{eq:d2n}) completes the proof.

\begin{proposition}
	\label{prop:loss}
	The loss function for training the proposed CNN refiner in (\ref{eq:DisCro_Mapping}) is bounded by its patch-based training loss function: 
	\begin{small}
		\begin{equation}
		\label{eq:conn}
		\centering
		\begin{split}
		\frac{1}{2L} \sum_{l=1}^{L} \left\|
		\left[ \hspace{-3pt} \begin{array}{c}
		\x_{l,1}\\ \x_{l,2}
		\end{array}  \hspace{-3pt} \right]   \hspace{-3pt}  -
		\left[    \hspace{-3pt}
		\begin{array}{c}
		\sum_{k=1}^{K}\sum_{n=1}^{2} \dd_{1,n,k} *\TT_{\exp{(\alpha_{n,k})}} \left( \sum_{m=1}^{2} \ee_{n,m,k} * \x_{l,m}^{(i-1)} \right) \\
		\sum_{k=1}^{K}\sum_{n=1}^{2} \dd_{2,n,k} *\TT_{\exp{(\alpha_{n,k})}} \left( \sum_{m=1}^{2} \ee_{n,m,k} * \x_{l,m}^{(i-1)} \right)
		\end{array}    \hspace{-3pt}
		\right]\right\|^2_2  \hspace{-4pt} \le \\
		\frac{1}{2LR} \sum_{l=1}^{L} \left\| 
		\left[   \hspace{-4pt}
		\begin{array}{c}
		\widetilde{\X}_{l,1}\\ \widetilde{\X}_{l,2}
		\end{array}  \hspace{-4pt}
		\right]   \hspace{-3pt}    - 
		\D\TT_{\exp{(\alp)}}\left(\hspace{-1pt} \E \left[ \hspace{-3pt} \begin{array}{c}
		\widetilde{\X}_{l,1}^{(i-1)} \\ \widetilde{\X}_{l,2}^{(i-1)}
		\end{array} \hspace{-3pt} \right]  \hspace{-1pt} \right)
		\right\|_{F}^2, 
		\end{split}
		\end{equation}   
	\end{small}
	where $\x_{l,m}$ and $\x_{l,m}^{(i-1)}$ are the $l$th high-quality and degraded images of the $m$th material, respectively, for $l=1,\dots,L$ and $m=1,2$, $\widetilde{\X}_{l,m}\in \mathbb{R}^{R\times N}$ and $\widetilde{\X}_{l,m}^{(i-1)} \in \mathbb{R}^{R\times N}$ are matrices whose columns are vectorized patches extracted from images $\x_{l,m}$ and $\x_{l,m}^{(i-1)}$ (with a spatial patch stride of $1\times 1$), respectively.
	See related notations in (\ref{eq:DisCro_Mapping}), (\ref{eq:conv2patch}), and Section~\ref{sec:Traing}.
\end{proposition}

\vspace{0.5pc}
\noindent\textit{Proof.} 
Based on Proposition~\ref{prop:refiner}, we obtain the result as follows:
\begin{footnotesize}
	\begin{equation*}
	\centering
	\begin{split}
	&\frac{1}{2L} \sum_{l=1}^{L} \left\|
	\left[ \hspace{-3pt} \begin{array}{c}
	\x_{l,1}\\\x_{l,2}
	\end{array} \hspace{-3pt} \right] -
	\left[
	\begin{array}{c}
	\sum_{k=1}^{K}\sum_{n=1}^{2} \dd_{1,n,k} *\TT_{\exp{(\alpha_{n,k})}} \left( \sum_{m=1}^{2} \ee_{n,m,k} * \x_{l,m}^{(i-1)} \right) \\
	\sum_{k=1}^{K}\sum_{n=1}^{2} \dd_{2,n,k} *\TT_{\exp{(\alpha_{n,k})}} \left( \sum_{m=1}^{2} \ee_{n,m,k} * \x_{l,m}^{(i-1)} \right)
	\end{array}
	\right]\right\|^2_2 \\
	=&\frac{1}{2L} \sum_{l=1}^{L} \left\|
	\left[ \hspace{-3pt} \begin{array}{c}
	\x_{l,1}\\ \x_{l,2}
	\end{array} \hspace{-3pt} \right]\hspace{-2pt}-\hspace{-2pt} \right.\\ &\left.
	\frac{1}{R}  \hspace{-3pt} \left[ \hspace{-3pt}\begin{array}{c}
	\sum_{j=1}^{N} \P_j^\top \D_{1,1} \TT_{\exp{(\alp_1)}} \left( \E_{1,1}\P_j\x_{l,1}^{(i-1)}\hspace{-2pt}+\hspace{-2pt}\E_{1,2}\P_j\x_{l,2}^{(i-1)}\right) \hspace{-3pt}+\hspace{-3pt} \sum_{j=1}^{N} \P_j^\top \D_{1,2} \TT_{\exp{(\alp_2)}} \left( \E_{2,1}\P_j\x_{l,1}^{(i-1)} \hspace{-2pt}+\hspace{-2pt} \E_{2,2}\P_j\x_{l,2}^{(i-1)}\right) \\
	\sum_{j=1}^{N} \P_j^\top \D_{2,1} \TT_{\exp{(\alp_1)}} \left( \E_{1,1}\P_j\x_{l,1}^{(i-1)}\hspace{-2pt}+\hspace{-2pt}\E_{1,2}\P_j\x_{l,2}^{(i-1)}\right)\hspace{-3pt}+\hspace{-3pt}\sum_{j=1}^{N} \P_j^\top \D_{2,2} \TT_{\exp{(\alp_2)}} \left( \E_{2,1}\P_j\x_{l,1}^{(i-1)}\hspace{-2pt}+\hspace{-2pt}\E_{2,2}\P_j\x_{l,2}^{(i-1)}\right)
	\end{array} \hspace{-3pt}\right]\right\|^2_2  \\
	=&\frac{1}{2LR^2} \sum_{l=1}^{L} \left\|
	\left[  \hspace{-5pt}  \begin{array}{c}
	\sum_{j=1}^{N}\P_j^\top\P_j\x_{l,1}\\\sum_{j=1}^{N}\P_j^\top\P_j\x_{l,2}
	\end{array}  \hspace{-5pt} \right] \hspace{-3pt} -\hspace{-3pt} \right.\\ &\left.
	\left[  \hspace{-5pt}  \begin{array}{c}
	\sum_{j=1}^{N} \P_j^\top \left(\D_{1,1} \TT_{\exp{(\alp_1)}} \hspace{-3pt}  \left( \E_{1,1}\P_j\x_{l,1}^{(i-1)} \hspace{-3pt} +\hspace{-3pt}  \E_{1,2}\P_j\x_{l,2}^{(i-1)}\right) \hspace{-3pt} +\hspace{-3pt}   \D_{1,2} \TT_{\exp{(\alp_2)}} \hspace{-3pt}  \left( \E_{2,1}\P_j\x_{l,1}^{(i-1)}\hspace{-3pt} +\hspace{-3pt} \E_{2,2}\P_j\x_{l,2}^{(i-1)}\right) \right)\\
	\sum_{j=1}^{N} \P_j^\top \left(\D_{2,1} \TT_{\exp{(\alp_1)}} \hspace{-3pt} \left( \E_{1,1}\P_j\x_{l,1}^{(i-1)}\hspace{-3pt} +\hspace{-3pt} \E_{1,2}\P_j\x_{l,2}^{(i-1)}\right)\hspace{-3pt} +\hspace{-3pt}  \D_{2,2} \TT_{\exp{(\alp_2)}}\hspace{-3pt}  \left( \E_{2,1}\P_j\x_{l,1}^{(i-1)}\hspace{-3pt} +\hspace{-3pt} \E_{2,2}\P_j\x_{l,2}^{(i-1)}\right)\right)
	\end{array}     \hspace{-5pt} \right]
	\right\|^2_2 \\
	\le& \frac{1}{2LR} \sum_{l=1}^{L}\sum_{j=1}^{N}\left\|
	\left[
	\begin{array}{c}
	\widetilde{\X}_{l,1,j} \\ \widetilde{\X}_{l,2,j}
	\end{array}
	\right]-
	\left[
	\begin{array}{cc}
	\D_{1,1} & \D_{1,2} \\
	\D_{2,1} & \D_{2,2}
	\end{array}
	\right]
	\TT_{\exp{(\alp)}}\left(
	\left[
	\begin{array}{cc}
	\E_{1,1} & \E_{1,2} \\
	\E_{2,1} & \E_{2,2}
	\end{array}
	\right]
	\left[
	\begin{array}{c}
	\widetilde{\X}_{l,1,j}^{(i-1)} \\ \widetilde{\X}_{l,2,j}^{(i-1)}
	\end{array}
	\right]\right)
	\right\|^2_2 \\
	=& \frac{1}{2LR} \sum_{l=1}^{L} \left\| 
	\left[
	\begin{array}{c}
	\widetilde{\X}_{l,1}\\ \widetilde{\X}_{l,2}
	\end{array}
	\right] - 
	\D\TT_{\exp{(\alp)}}\left(\E \left[ \begin{array}{c}
	\widetilde{\X}_{l,1}^{(i-1)} \\ \widetilde{\X}_{l,2}^{(i-1)}
	\end{array}\right] \right)
	\right\|_{F}^2,
	\end{split} 
	\end{equation*}
\end{footnotesize}
where $\widetilde{\X}_{l,m,j}^{(i-1)}\in\mathbb{R}^R$ and $\widetilde{\X}_{l,m,j}\in\mathbb{R}^R$ are the $j$th column of $\widetilde{\X}_{l,m}^{(i-1)}$ and $\widetilde{\X}_{l,m}$, respectively.
Here, the inequality holds by  $\widetilde{\P}\widetilde{\P}^\top \preceq R\cdot \I$ with $\widetilde{\P}:={[\P_1^\top, \cdots, \P_{N}^\top]}^\top$.

\section{Supplementary Results for Section~\ref{sec:experiments}}
\label{sec:result-supp}

\begin{figure}[!t]
	\centering
	\includegraphics[scale=1.37]{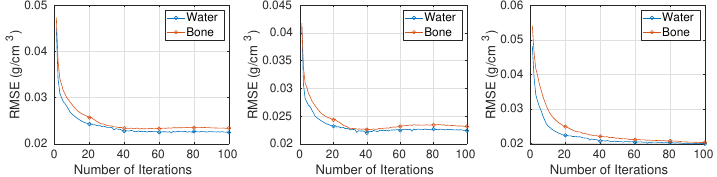}
	\caption{RMSE plot of BCD-Net-dCNN for Test \#1, Test \#2, and Test \#3, respectively.}
	\label{fig:rmse-plot-10epoch}
\end{figure}

Figure~\ref{fig:rmse-plot-10epoch} shows the RMSE plots of water and bone images for BCD-Net-dCNN.
BCD-Net-dCNN becomes overfitted around 40th iteration for test slices \#1 and \#2.

\begin{figure}[!t]
	\vspace{6pt}
	\centering
	\includegraphics[scale=1.37]{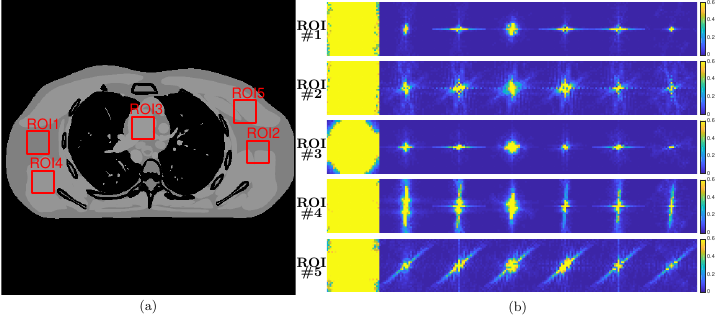}
	\vspace{-10pt}
	\caption{(a) Five selected ROIs indicated for $\overline{\textup{NPS}}$ calculation for the decomposed water image of XCAT phantom. (b) Left to right: NPS measured within ROIs of decomposed water images obtained by direct matrix inversion, DECT-EP, DECT-ST, dCNN, BCD-Net-dCNN, BCD-Net-sCNN-lc, and BCD-Net-sCNN-hc.
			The first to the fifth rows in (b) show the $\overline{\textup{NPS}}$ of the first to fifth ROIs, respectively, with display windows [0 0.6]~g$^2$/cm$^6$.}
	\label{fig:nps-xcat}
\end{figure}

We generated ten different noise realizations to obtain NPS images for XCAT phantom data.
We calculated the averaged NPS measure~\cite{supp-nps:14}, denoted as $\overline{\textup{NPS}}$, for each method using
\begin{equation*}
\overline{\textup{NPS}}=\frac{\sum_{i=1}^{10} |\mathrm{DFT}\{f_i-f^*\}|^2}{10},
\end{equation*} 
where $f_i$ denotes the decomposed water image from the $i$th noise realization, and $f^*$ denotes the ground truth of water image.
Figure~\ref{fig:nps-xcat} compares the magnitude of $\overline{\textup{NPS}}$ from different methods.
Across all frequencies, the NPS magnitude of BCD-Net-sCNN-hc is significantly smaller than those of direct matrix inversion, DECT-EP, DECT-ST, and dCNN.
Furthermore, BCD-Net-sCNN-hc gives fewer vertical and horizontal frequency strips with lower intensity, compared to BCD-Net-sCNN-lc and BCD-Net-dCNN.
The aforementioned NPS comparisons demonstrate the superiority of the proposed BCD-Net-sCNN-hc method in removing noise and artifacts inside soft tissue regions.

\begin{figure*}[!t] 
	\centering
	\includegraphics[scale=1.37]{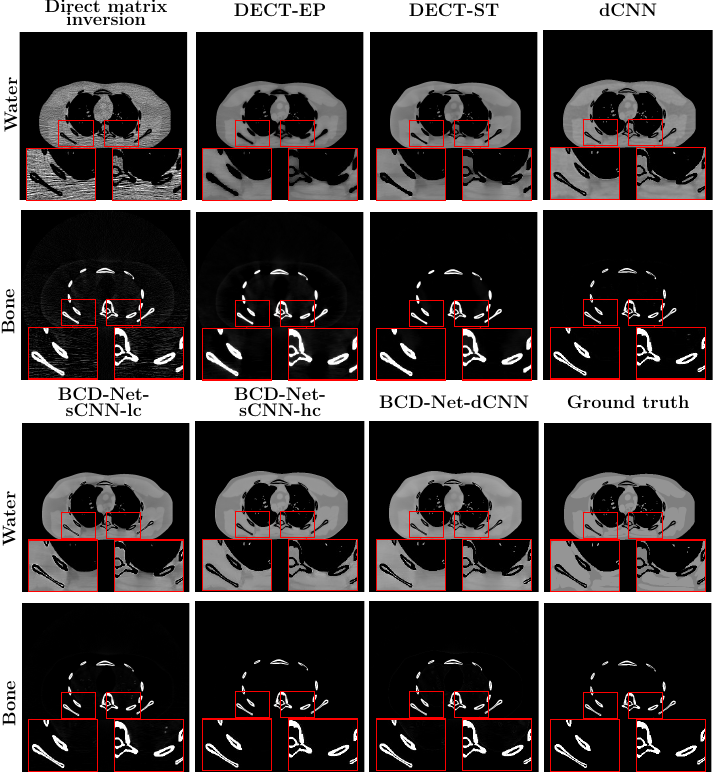}
	\vspace{-6pt}
	\caption{Comparison of decomposed images from different methods (XCAT phantom test slice \#2). Water and bone images are shown with display windows [0.7  1.3]\,g/cm$^3$ and [0 0.8]\,g/cm$^3$, respectively.}		
	\label{fig:test2}
\end{figure*}

\begin{figure*}[!t] 
	\centering
	\includegraphics[scale=1.37]{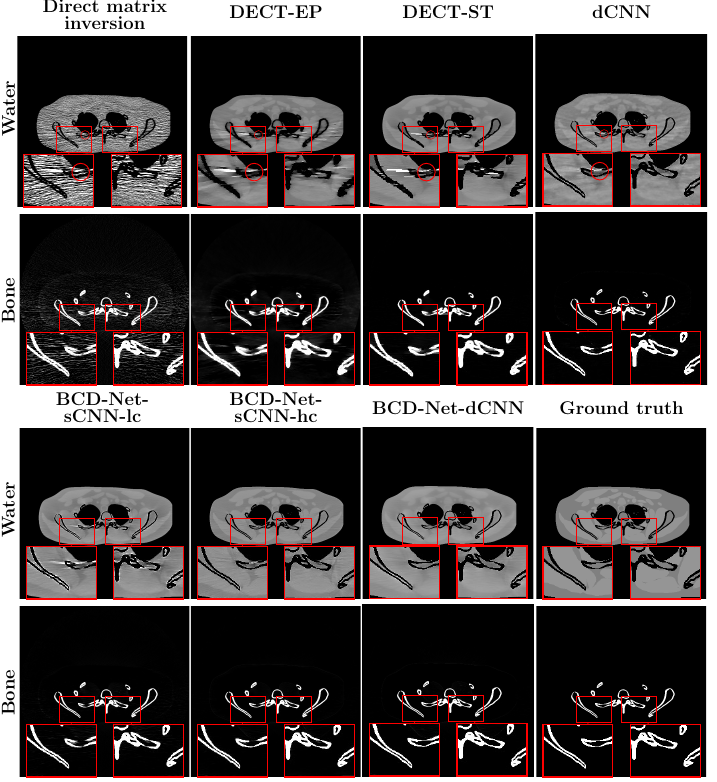}
	\vspace{-10pt}
	\caption{Comparison of decomposed images from different methods (XCAT phantom test slice \#3).  
		Water and bone images are displayed with windows [0.7  1.3]\,g/cm$^3$ and [0 0.8]\,g/cm$^3$, respectively.}
	\label{fig:test3}
	\vspace{-5pt}
\end{figure*}

Figure~\ref{fig:test2} and Figure~\ref{fig:test3} show another two test slices comparisons.
DCNN improves decomposition quality compared to DECT-EP and DECT-ST in terms of reducing noise and artifacts, but it still retains some streak artifacts. 
Compared to DCNN, BCD-Net-sCNN-hc further removes noise and artifacts, and improves the sharpness of edges in soft tissue.

\clearpage

\begin{figure*}[!t] 
	\centering
	\includegraphics[scale=1.37]{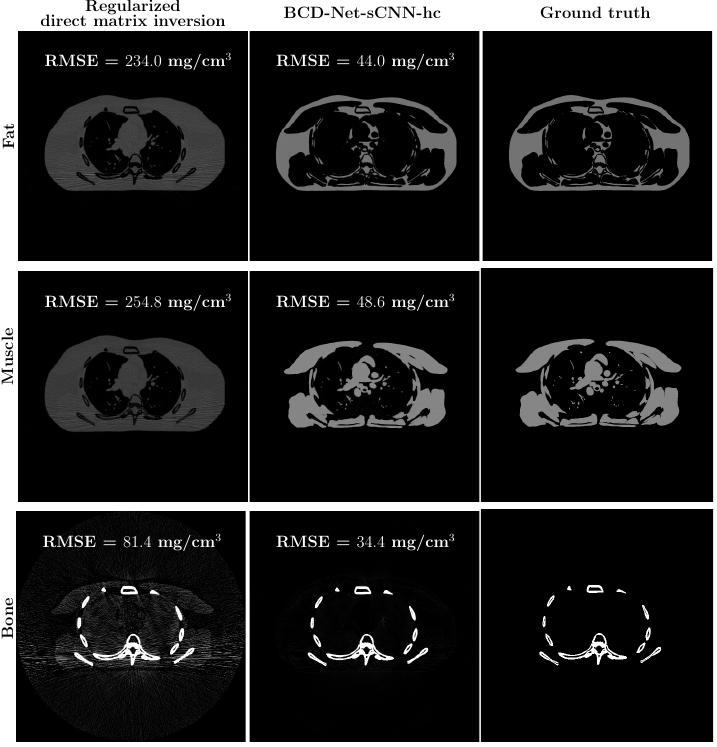}
	\caption{Comparison of three decomposed images from regularized direct matrix inversion ($\lambda=1\times 10^{-5}$), BCD-Net-sCNN-hc, and ground truth. 
		Fat, muscle, and bone images are shown with display windows [0 2]~g/cm$^3$, [0 2]~g/cm$^3$, and [0 0.5]~g/cm$^3$, respectively.}
	\label{fig:mmd}
\end{figure*}

\begin{figure}[!t]
	\centering
	\includegraphics[scale=0.6]{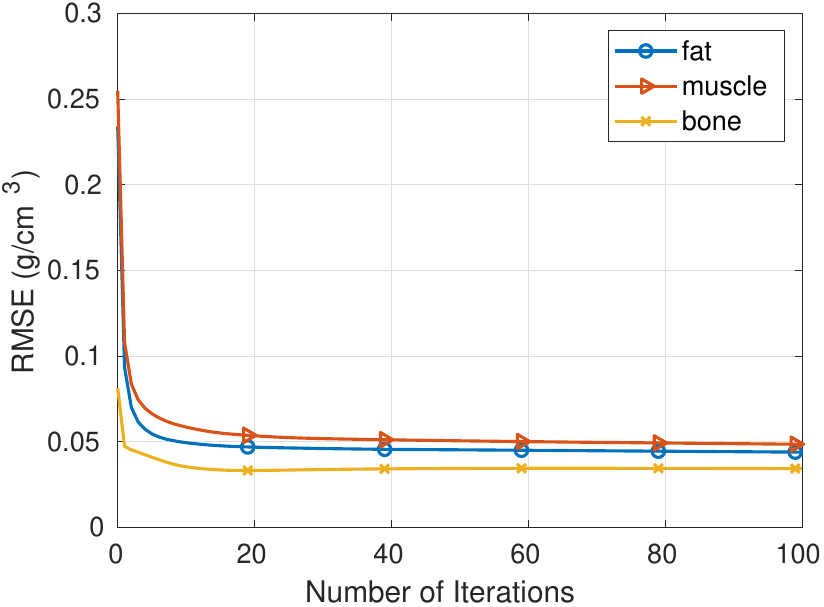}
	\vspace{-4pt}
	\caption{RMSE convergence behaviors of three-material decomposition BCD-Net-sCNN-hc.}
	\label{fig:mmd_rmse}
	\vspace{-4pt}
\end{figure}

\begin{figure*}[!t] 
	\centering
	\includegraphics[scale=1.43]{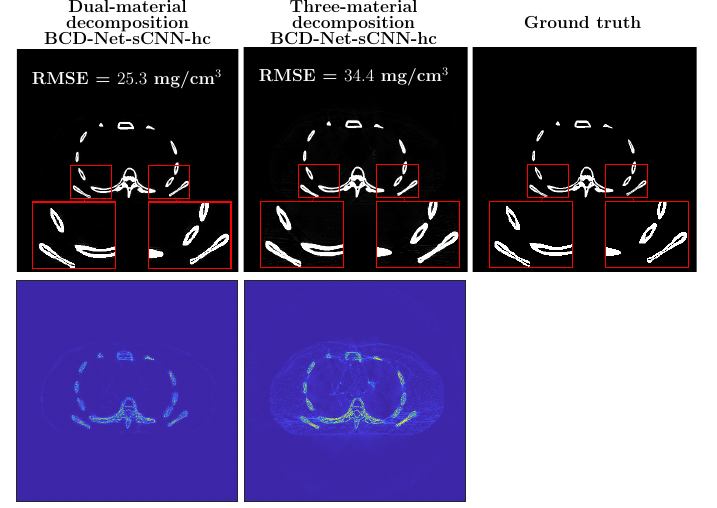}
	\caption{Comparisons of decomposed bone images (display window [0 0.5]~g/cm$^3$) and their error maps (display window [0 0.3]~g/cm$^3$) from dual- and three-material decomposition BCD-Net-sCNN-hc architectures.}
	\label{fig:comp-two-three}
\end{figure*}

	We ran additional three-material (fat, muscle, and bone) decomposition experiments with the proposed architecture, BCD-Net-sCNN-hc.
	We obtained the three initial decomposed images from high- and low-energy attenuation images, by using a Tikhonov-regularized direct matrix inversion method, i.e., $\x^{(0)}=(\A'\A + \lambda \I)^{-1} \A'\y$ (three-material decomposition in dual-energy CT is an under-determined inverse problem).
	Figure~\ref{fig:mmd} compares \#1 material density images from regularized direct matrix inversion,  BCD-Net-sCNN-hc, and ground truth.
	The regularized direct matrix inversion method suffers from severe noise and artifacts, and does not decompose fat and muscle images.
	BCD-Net-sCNN-hc achieves significantly better three-material decomposition performance over the regularized direct matrix inversion method.
	Figure~\ref{fig:mmd_rmse} shows the RMSE convergence behavior of BCD-Net-sCNN-hc: similar to the RMSE convergence behavior in dual-material decomposition (see Figure~\ref{fig:average_rmse_plot}), it decreases monotonically.
	Figure~\ref{fig:comp-two-three} compares decomposed bone images and their error maps from dual- and three-material decomposition BCD-Net-sCNN-hc.
	(Note that ground-truth bone images are identical between the dual- and three-material decomposition cases.)
	The dual-material decomposition BCD-Net architecture achieves smaller errors and clearer image edges and structures, compared to the three-material decomposition BCD-Net method; see error maps and zoom-ins in bone images.
	This is natural because the initial decomposed images from the dual-material decomposition case are more accurate than those from the three-material decomposition case, and $\A_0$ in (\ref{eq:MBID}) in dual-material decomposition is better conditioned than the counterpart in three-material decomposition in DECT.

\section*{References}
\addcontentsline{toc}{section}{\numberline{}References}
\vspace*{-20mm}
\bibliographystyle{./medphy.bst} 
\bibliography{./refs_bcdnet}      
\makeatletter\@input{xx.tex}\makeatother